\documentclass[aps,prd,preprint,nofootinbib,amsmath,amssymb]{revtex4}

\usepackage{amssymb}
\usepackage{mathrsfs}
\usepackage{slashed}
\usepackage{graphicx,color}
\usepackage{bm}
\usepackage{times}
\usepackage{amsmath}
\usepackage{subeqnarray}
\usepackage{multirow}

\usepackage{epsfig}
\usepackage{dcolumn}
\usepackage{float}
\usepackage{hyperref}
\hypersetup{
  colorlinks=true,
  citecolor=blue,
  linkcolor=blue,
  urlcolor=blue}

\newcommand{\dcp}{\delta_{CP}}
\newcommand{\nova}{NO$\nu$A\ }
\newcommand{\cnv}{\v{C}erenkov}

\begin{document}

\renewcommand{\arraystretch}{1.3}

\author{Shinya Fukasawa}
\email[e-mail: ]{fukasawa-shinya@ed.tmu.ac.jp}
\affiliation{
Department of Physics, Tokyo Metropolitan University, Hachioji, Tokyo 192-0397, Japan}

\author{Monojit Ghosh}
\email[e-mail: ]{monojt@tmu.ac.jp}
\affiliation{
Department of Physics, Tokyo Metropolitan University, Hachioji, Tokyo 192-0397, Japan}

\author{Osamu Yasuda}
\email[e-mail: ]{yasuda@phys.se.tmu.ac.jp}
\affiliation{
Department of Physics, Tokyo Metropolitan University, Hachioji, Tokyo 192-0397, Japan}

\title{Complementarity Between Hyperkamiokande and DUNE in Determining Neutrino Oscillation Parameters}

\begin{abstract}

In this work we investigate the sensitivity to the neutrino mass hierarchy, the octant of the mixing angle $\theta_{23}$
and the CP phase $\dcp$ in the future long baseline
experiments T2HK and DUNE as well as in the atmospheric neutrino
observation at Hyperkamiokande (HK).
We show for the first time that the sensitivity is enhanced greatly if we combine these three experiments.
Our results show that the hierarchy sensitivity of both T2HK and HK are limited due to the presence of parameter degeneracy. But this degeneracy is removed when 
T2HK and HK are added together.
With T2HK+HK (DUNE), the neutrino mass hierarchy can be determined at least at
$ 5 \sigma$ (8\,$\sigma$) C.L. for any value of true $\dcp$.
With T2HK+HK+DUNE
the significance of the mass hierarchy increases to almost 15 $\sigma$ for the unfavorable value of $\dcp$.
For these combined setup, octant can be resolved except  $43.5^\circ < \theta_{23} < 48^\circ$ at $5\sigma$ C.L for both the hierarchies irrespective of the value of $\dcp$.
The significance of CP violation is around 10\,$\sigma$ C.L. for $\dcp\sim \pm 90^\circ$. Apart from that these combined facility has the capability to discover CP violation
for at least $68\%$ fraction of the true $\dcp$ values at $5 \sigma$ for any value of true $\theta_{23}$. 
We also find that, with combination of all these three, the precision of $\Delta m^2_{{\rm eff}}$, $\sin^2\theta_{23}$ and $\dcp$ becomes 0.3\%, 2\% and 20\% respectively.
We also clarify how the octant degeneracy occurs in the HK atmospheric neutrino experiment.

\end{abstract}

\maketitle

\section{Introduction}
With the discovery of the mixing angle $\theta_{13}$ by the 
reactor experiments \cite{An:2015rpe,dchooz_latest,RENO:2015ksa}, 
the physics of neutrino oscillation has entered into an era of precision measurement. In the standard three flavor scenario, the phenomenon of 
neutrino oscillation can be parametrized by three mixing angles:
$\theta_{12}$, $\theta_{23}$ and $\theta_{13}$, two mass squared difference: $\Delta m_{21}^2$ and $\Delta m_{31}^2$, and one Dirac type phase $\dcp$.
Thanks to the neutrino experiments in the last two decades,
the values of the three mixing angles and the values of the mass squared differences are now determined
in the three flavor mixing framework to some precision \cite{Capozzi:2013csa,Forero:2014bxa,Bergstrom:2015rba}.
The unknown quantities at present are: (i) the mass hierarchy or the sign of $\Delta m_{31}^2$ (NH: normal hierarchy i.e., $\Delta m_{31}^2 > 0$ or IH: 
inverted hierarchy i.e., $\Delta m_{31}^2 < 0$ ),
(ii) the octant of $\theta_{23}$ (LO: lower octant i.e., $\theta_{23} < 45^\circ$ or HO: higher octant i.e., $\theta_{23} > 45^\circ$ ) and 
(iii) the CP phase $\dcp$. There are several experiments which are dedicated to 
determine these above mentioned unknowns.

The main difficulty in determining those unknowns is the presence of parameter degeneracy \cite{BurguetCastell:2001ez,Minakata:2001qm,Fogli:1996pv,Barger:2001yr,Ghosh:2015ena}. In parameter degeneracy, different sets of oscillation
parameter leads to the same value of the oscillation probability. Due to this the true solutions can be mimicked by the false solutions and thus a unique 
determination of the parameters becomes difficult.
One of the ways to overcome this degeneracy is to combine data from different experiments. As the degenerate parameter space is different for different experiments, 
combination of different experiments can lead to the
removal of the fake solutions which may help in the unambiguous determination of the neutrino oscillation parameters. 
Recently, this strategy has been adopted by many to study the synergy between different
on-going as well as future proposed experiments in determining the remaining unknowns of the neutrino oscillation parameter.
In this regard, the well established method is to combine the data of the long-baseline and atmospheric neutrino oscillation experiments. 
The sensitivity to the remaining unknown parameters in long-baseline experiments comes from the appearance channel probabilities
$P(\nu_\mu \rightarrow \nu_e)$ and $P(\bar{\nu}_\mu \rightarrow \bar{\nu}_e)$ while the atmospheric experiments can have sensitivity for both appearance and disappearance channel 
probabilities $P(\nu_\mu \rightarrow \nu_\mu)$ and $P(\bar{\nu}_\mu \rightarrow \bar{\nu}_\mu)$. 
In addition to that some of the authors also prefer to add the reactor data which has the sensitivity of the electron antineutrino disappearance channel ($\bar{\nu}_e \rightarrow \bar{\nu}_e$).
Thus the combination of the long-baseline and atmospheric
experiments along with the information of the precise value of $\theta_{13}$, is believed to remove the degenerate solutions and measure the current unknowns at a significant confidence level.
In Refs. \cite{Ghosh:2013zna,Ghosh:2014dba,Ghosh:2015tan}, the combined analysis on the CP sensitivity of the long-baseline experiments T2K \cite{Abe:2014tzr}, \nova \cite{nova_tdr}
and atmospheric experiment ICAL@INO \cite{inowhitepaper} has been studied in detail. The combined analysis on the octant sensitivity of these experiments can be found in \cite{Chatterjee:2013qus}. 
In these analyses, the reactor information has been taken into account in the form of prior on $\sin^22\theta_{13}$.
The hierarchy sensitivity of T2K, \nova and ICAL@INO along with the combination of the reactor experiments has been done in \cite{Ghosh:2012px}.
The sensitivity of the long-baseline experiment LBNO \cite{Stahl:2012exa} with the addition of T2K, \nova and ICAL@INO has been studied in \cite{Ghosh:2013pfa}.
The combined analysis on the octant sensitivity of the atmospheric experiment PINGU \cite{Aartsen:2014oha} along with T2K, \nova and reactor data has been
studied in Ref. \cite{Choubey:2013xqa}. The synergistic study on the hierarchy sensitivity of PINGU by combining reactor experiment has been done in \cite{Blennow:2013vta}
whereas the same with combining beam based experiments and reactors is carried out in \cite{Winter:2013ema}.
The synergy of the proposed long-baseline experiment DUNE \cite{Acciarri:2015uup}, 
combining both the beam and atmospheric data along with T2K, \nova is analyzed in \cite{Barger:2013rha,Barger:2014dfa}.
The combined sensitivity of DUNE in conjunction with T2K, \nova and ICAL@INO has been explored in \cite{Ghosh:2014rna}. 
All these studies shows that, the sensitivity of an individual experiment is significantly enhanced when added with 
the other experiments.

In this paper, for the first time we study the joint sensitivity of the long-baseline experiments T2HK \cite{Abe:2014oxa}, DUNE and the atmospheric experiment HK \cite{hkloi} in determining the 
remaining unknowns in neutrino oscillation sector. This is the so-called HK and LBNF complementarity which has been discussed in the literature \cite{Cao:2015ita}.
We take this opportunity to work out the physics potential of these facilities in detail.
The T2HK experiment is an upgrade of the ongoing T2K experiment which will use a detector to have a volume
almost 25 times larger than the existing T2K detector. HK is the atmospheric counterpart of the T2HK experiment. On the other hand DUNE is a high statistics beam based
experiment to use high beam power, large detector volume and longer baseline. Among all other existing facilities, these above mentioned experiments are the most promising
future experiments in terms of both statistics and matter effect which have the maximum potential to reveal the true nature of the neutrino oscillation parameters.  
In this work we study: (i) the sensitivity of the T2HK, HK and DUNE experiments, (ii) the synergy between the T2HK and HK experiments to resolve the parameter degeneracy in the neutrino oscillation, 
(iii) how far the sensitivities in determining hierarchy, octant and CP can be stretched when all these three powerful experiments are combined together and 
(iv) the precision measurements of $\theta_{23}$, $\dcp$ and $\Delta m_{31}^2$ of this setup. 
 
The paper is organized as follows.
In Sect.\,\ref{preliminaries}, we
describe a little about parameter degeneracy,
which becomes important for determination of
the CP phase, and the experiments T2HK, DUNE
and the atmospheric neutrino measurement at
Hyperkamiokande (HK).
In Sect.\,\ref{analysis}, we give our simulation details
and the results of our analysis.
In Sect.\,\ref{conclusion}, we draw our conclusions.
In the appendix\,\ref{appendixa}, we provide a discussion regarding the 
parameter degeneracy in the HK atmospheric neutrino experiment.

\section{Preliminaries} 
\label{preliminaries}

\begin{figure*}
\begin{center}
\begin{tabular}{lr}
\hspace*{-0.1in}
\includegraphics[width=0.25\textwidth]{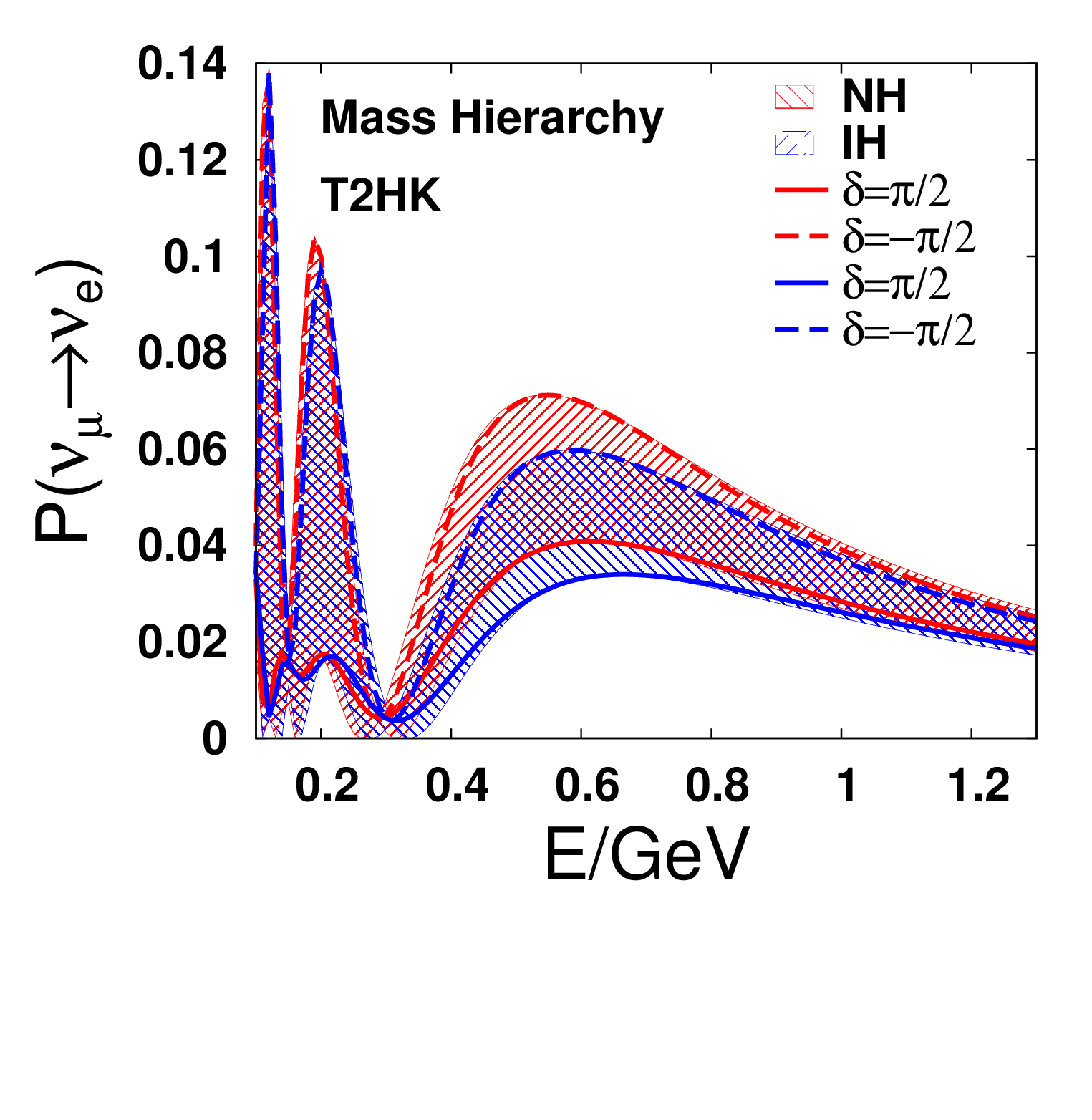} 
\hspace*{-0.1in}
\includegraphics[width=0.25\textwidth]{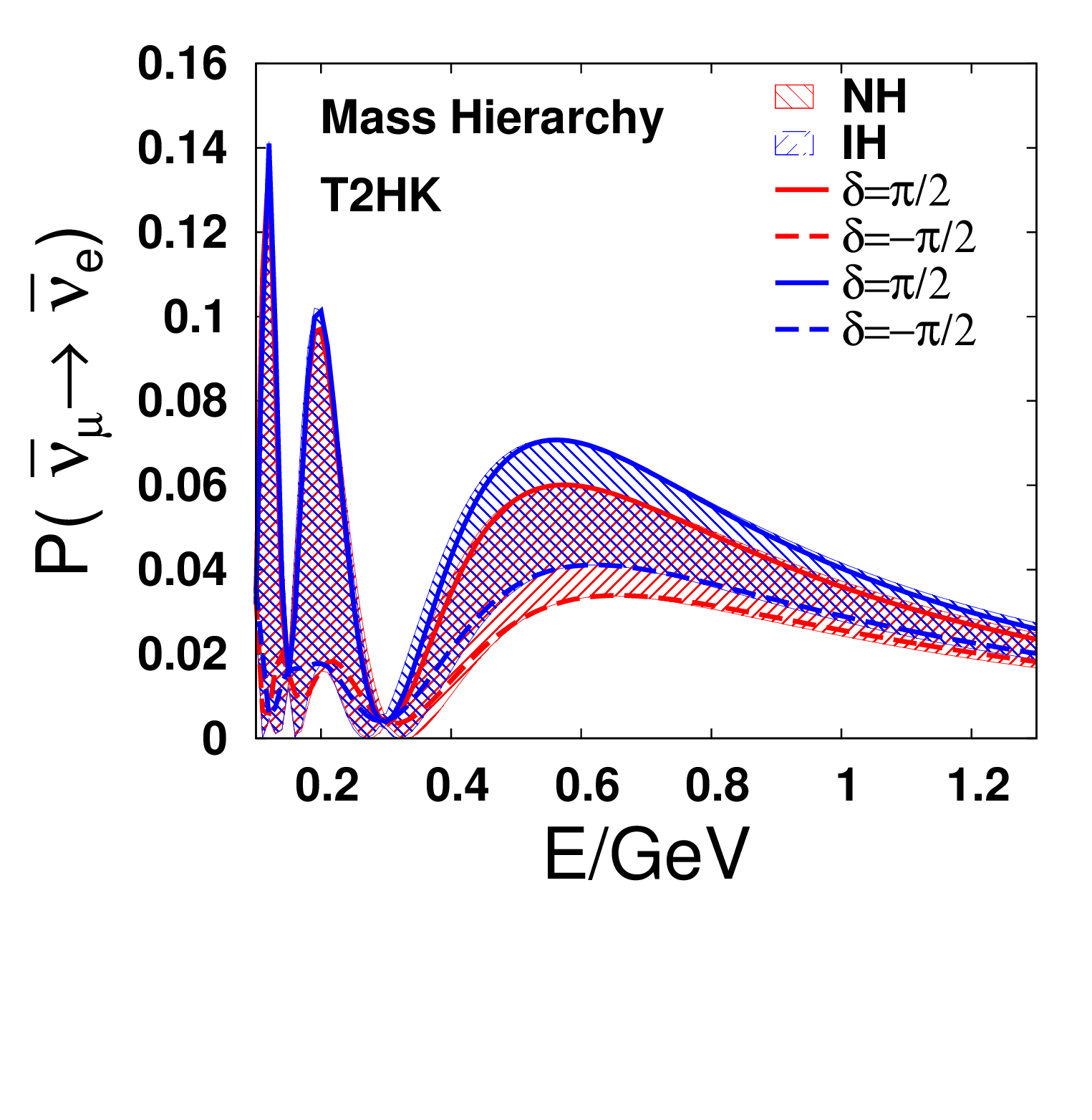}
\hspace*{-0.1in}
\includegraphics[width=0.25\textwidth]{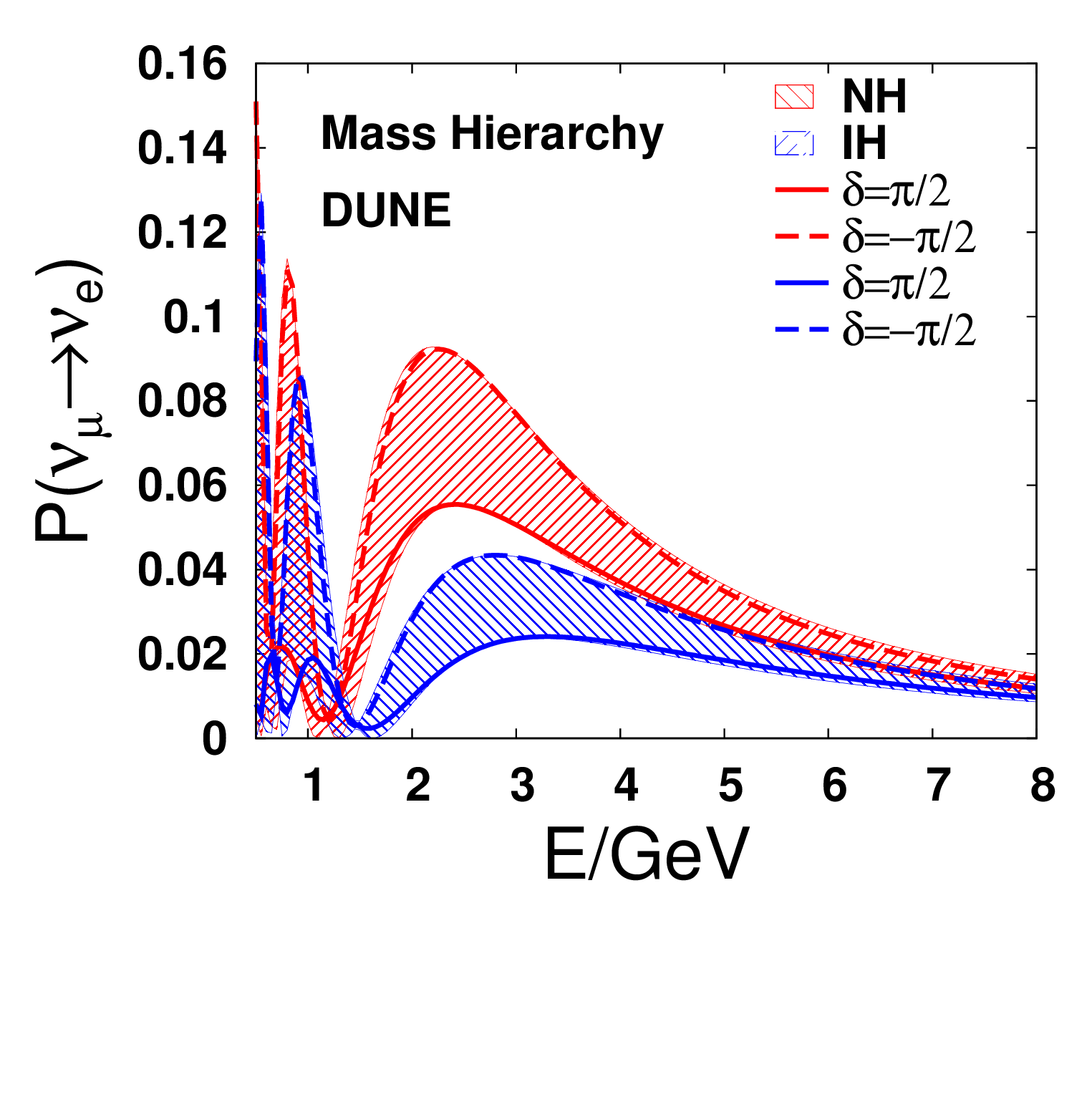}
\hspace{-0.1in}
\includegraphics[width=0.25\textwidth]{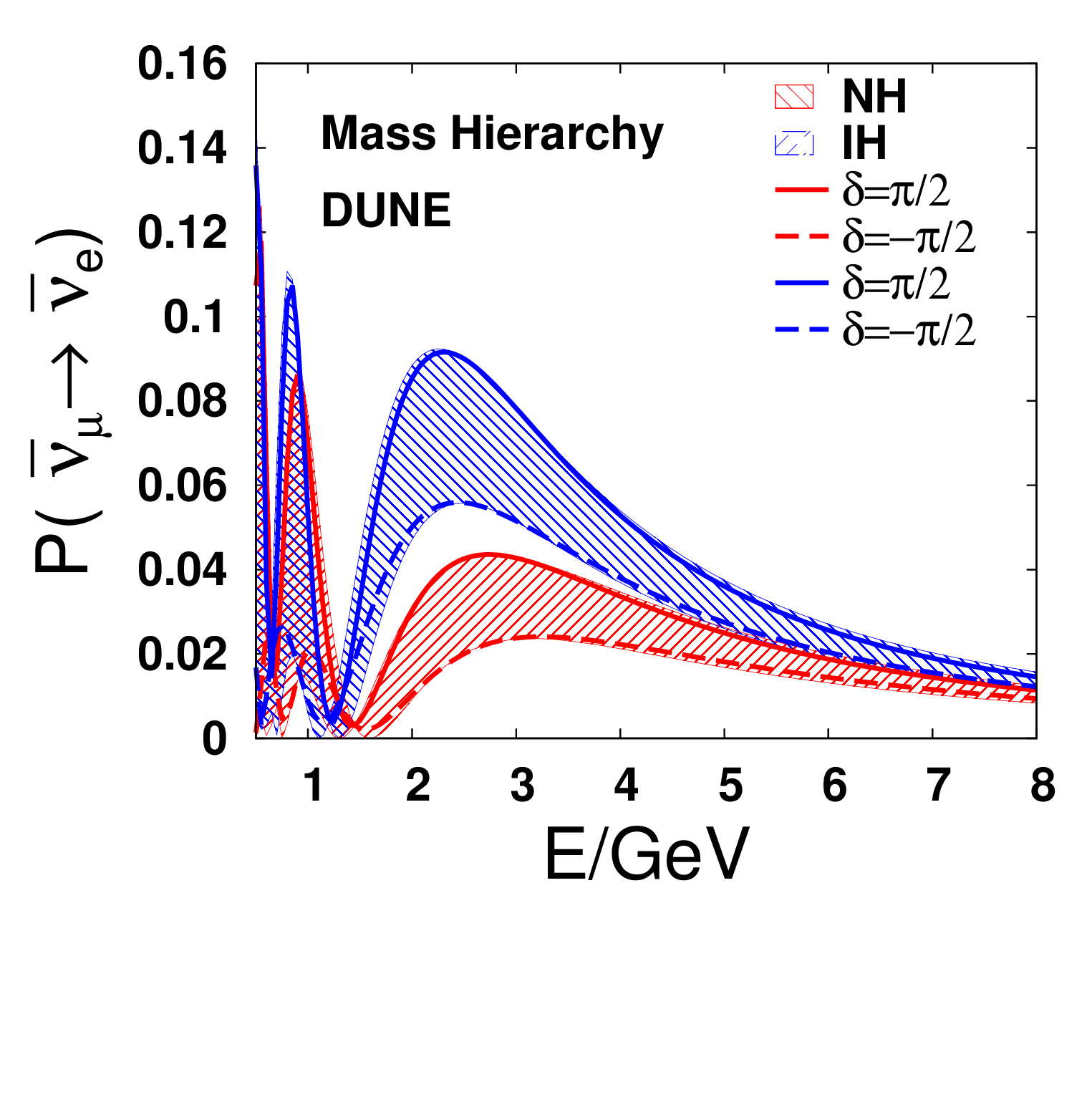}
\end{tabular}
\end{center}
\vspace{-1.5cm}
\caption{The behaviors of the appearance probabilities
$P(\nu_\mu\to\nu_e)$ (first and third panels) and
$P(\bar{\nu}_\mu\to\bar{\nu}_e)$ (second and fourth panels)
for T2HK (left two panels) and DUNE (right two panels) in the
two different mass hierarchies.
The width of each band comes from the
degree of freedom of $\dcp$, and the edge
is approximately represented by $\dcp=\pm\,90^\circ$.
$\theta_{23}=45^\circ$ is assumed in all the four figures.
}
\label{t2hk-dune-hierarchy}
\end{figure*}

\begin{figure*}
\begin{center}
\begin{tabular}{lr}
\hspace{-0.1in}
\includegraphics[width=0.25\textwidth]{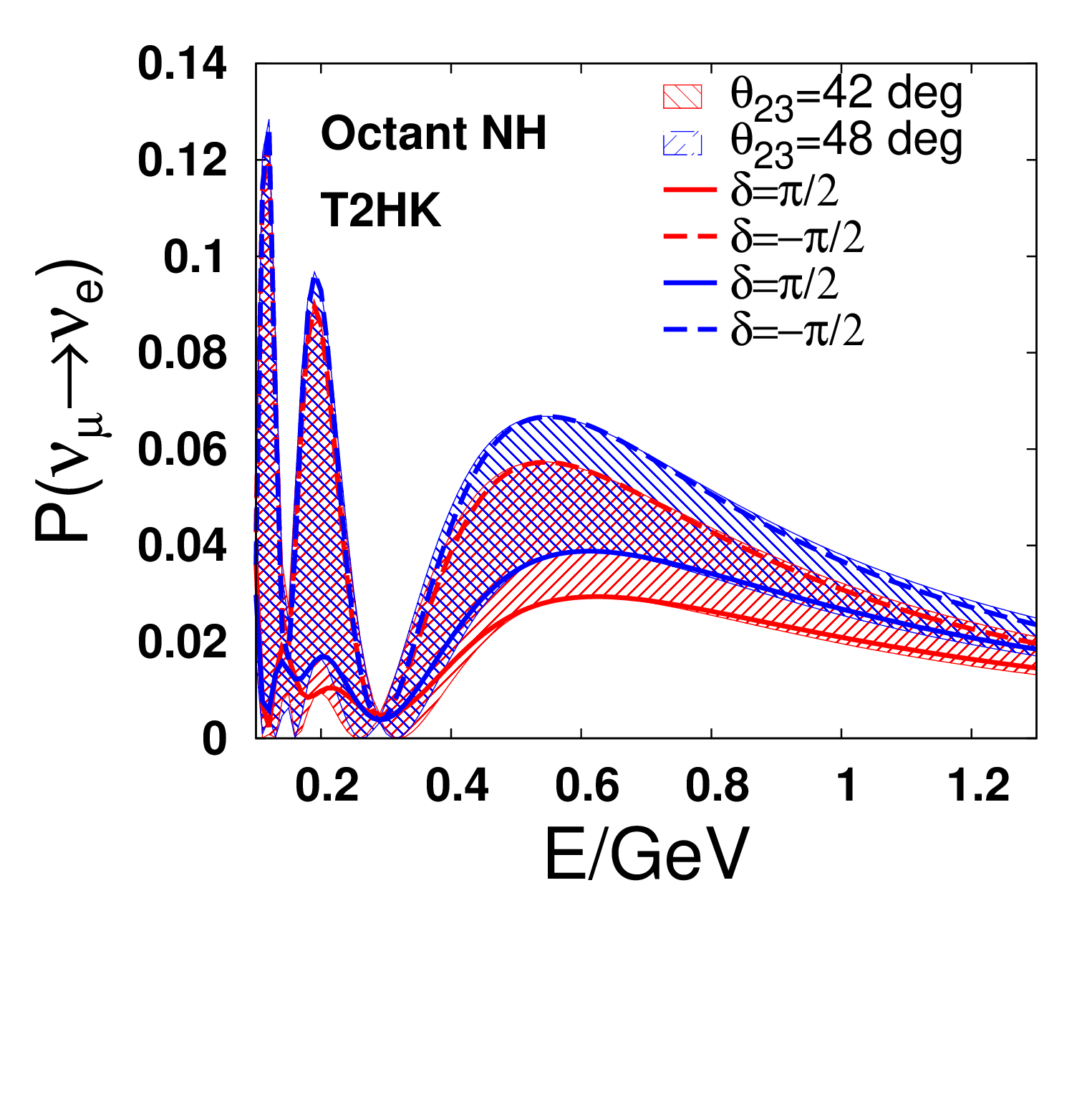} 
\hspace*{-0.1in}
\includegraphics[width=0.25\textwidth]{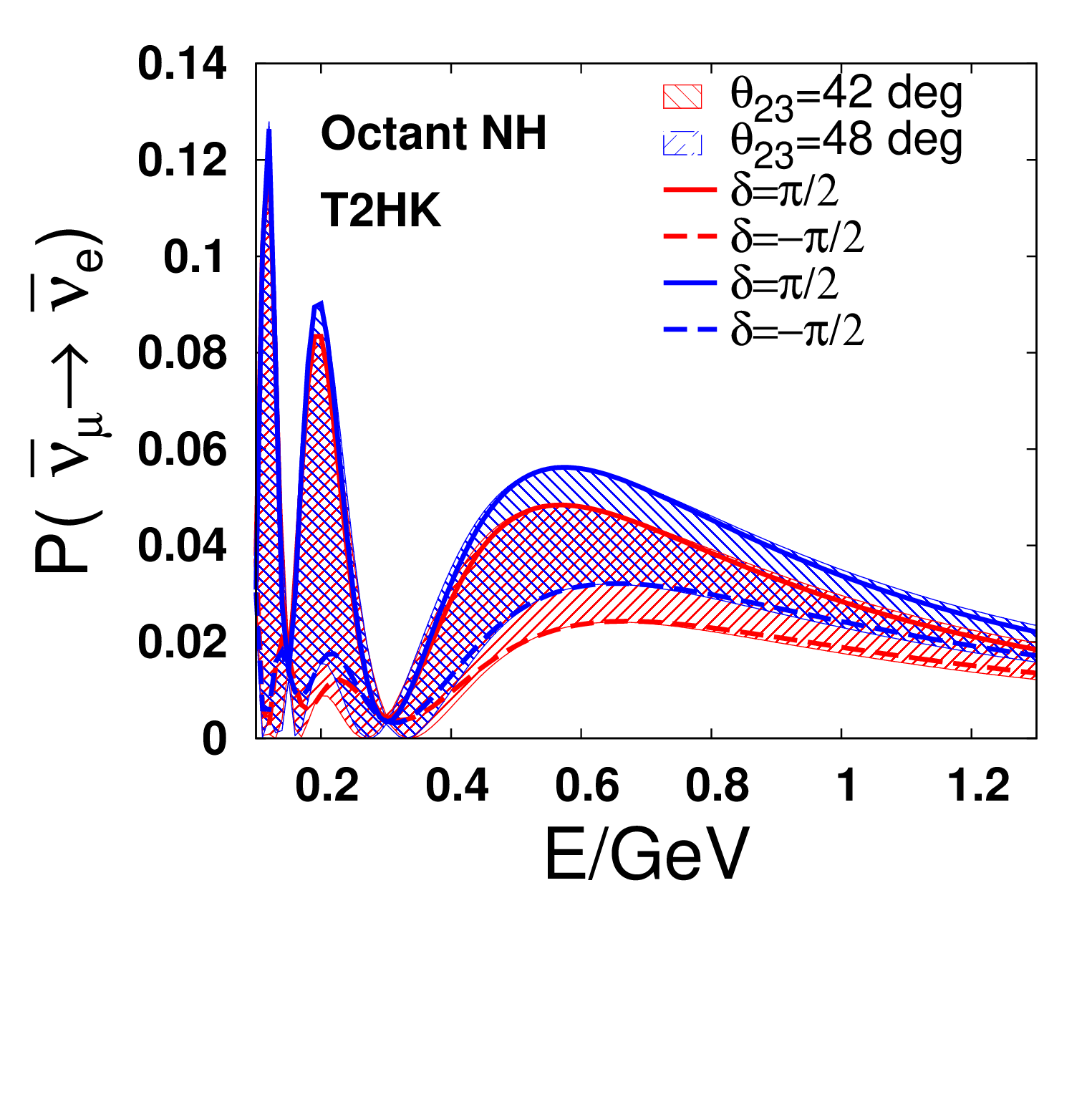}
\hspace{-0.1in}
\includegraphics[width=0.25\textwidth]{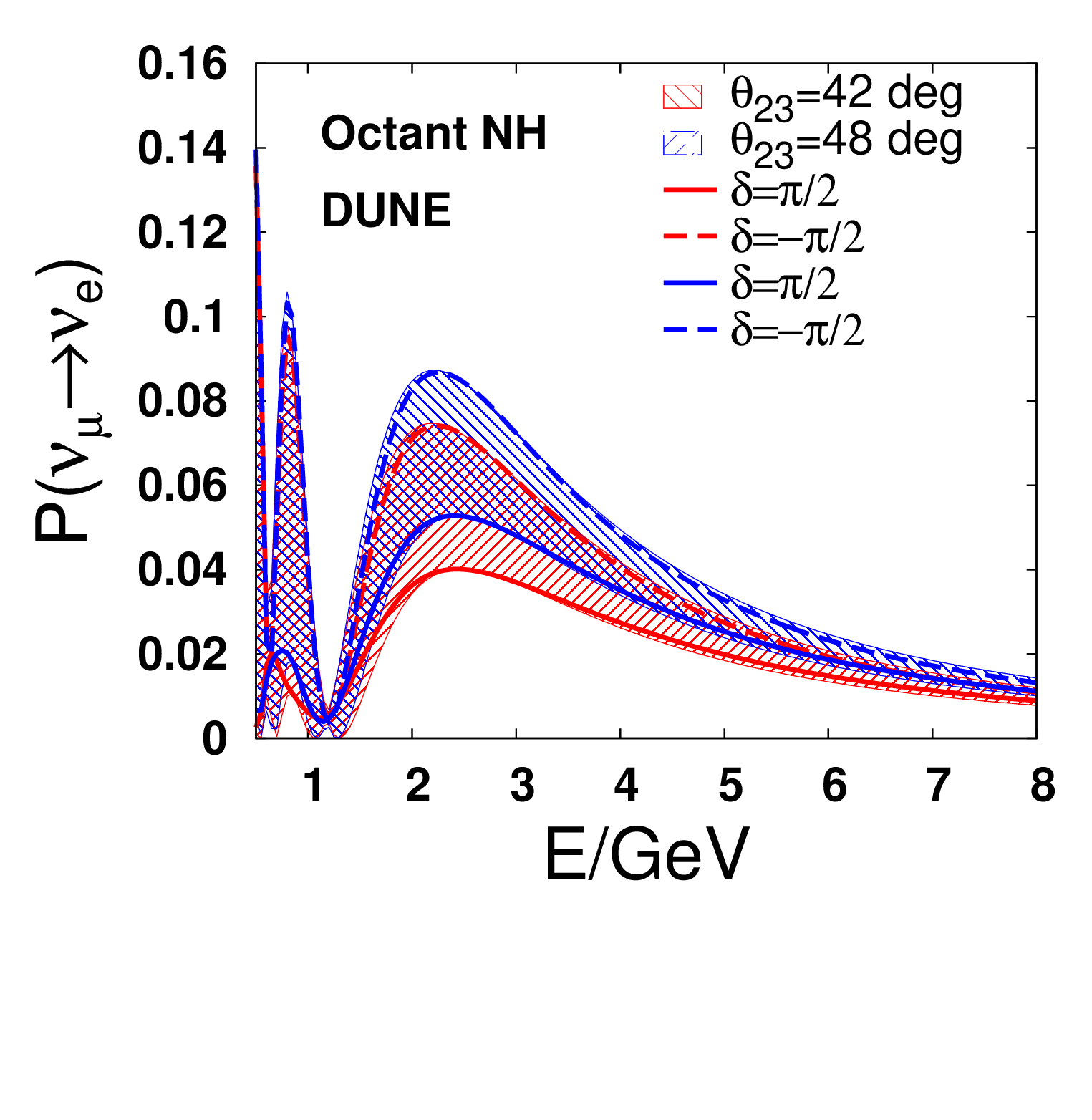}
\hspace{-0.1in}
\includegraphics[width=0.25\textwidth]{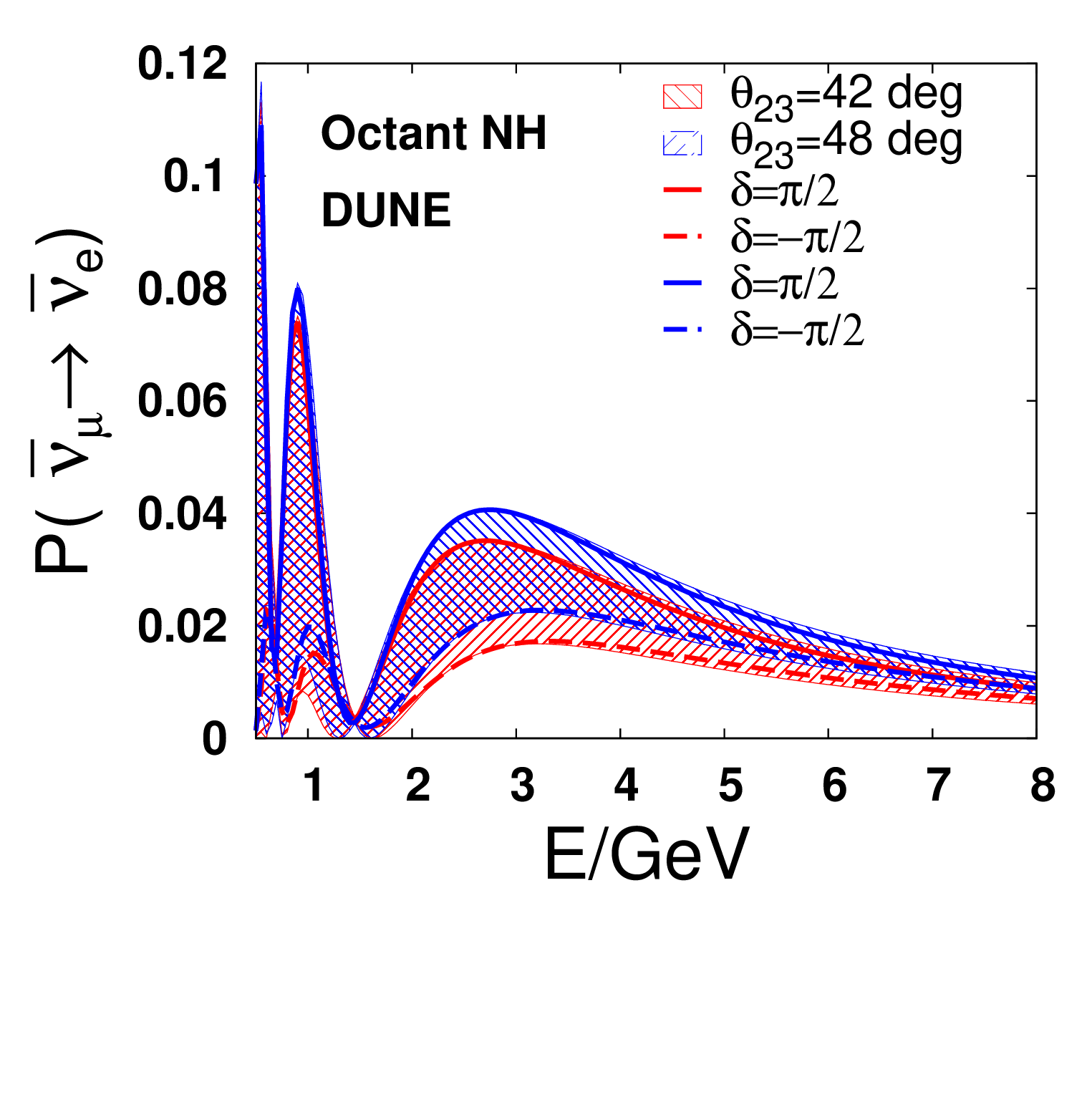}
\end{tabular}
\end{center}
\vspace{-1.5cm}        
\caption{The behaviors of the appearance probabilities
$P(\nu_\mu\to\nu_e)$ (first and third panels) and
$P(\bar{\nu}_\mu\to\bar{\nu}_e)$ (second and fourth panels)
for T2HK (left two panels) and DUNE (right two panels) in the
two different octants, where $\theta_{23}=42^\circ$ or
$\theta_{23}=48^\circ$ is taken as a reference value.
Normal hierarchy is assumed in all the four figures.
The width of each band comes from the
degree of freedom of $\dcp$, and the edge
is approximately represented by $\dcp=\pm\, 90^\circ$.
}
\label{t2hk-dune-octant}
\end{figure*}

\subsection{Parameter degeneracy}

The ultimate goal of the research on neutrino oscillations
in the standard three flavor mixing scheme is the
determination of $\dcp$, since it is expected to be relevant
to the baryon asymmetry of the universe.
It has been known that the determination of $\dcp$
is not easy because of so-called
parameter degeneracy:
Even if we know the
values of the appearance oscillation probabilities
$P(\nu_\mu\to\nu_e)$ and $P(\bar{\nu}_\mu\to\bar{\nu}_e)$
precisely,
we cannot determine the oscillation parameters
uniquely.
Earlier when $\theta_{13}$ was not known precisely,
there were three kinds of
parameter degeneracy.
The first parameter degeneracy is the intrinsic
degeneracy \cite{BurguetCastell:2001ez},
which occurs because the appearance oscillation probabilities
are approximately quadratic in $\sin2\theta_{13}$
for small $\theta_{13}$ and we obtain the
two solutions $(\theta_{13}, \dcp)$ and
$(\theta_{13}^\prime, \dcp^\prime)$. This intrinsic
degeneracy is not expected to cause a problem
because we now know the value of $\theta_{13}$
from the precise measurements of $\theta_{13}$
by the reactor neutrino experiments \cite{dchooz_latest,An:2015rpe,RENO:2015ksa}.
The second one is the sign degeneracy\,\cite{Minakata:2001qm}.
At this moment we do not know whether the mass hierarchy
is normal or inverted.
Depending on whether the mass hierarchy is normal or inverted,
the appearance oscillation probabilities vary, and
ignorance of the mass hierarchy may give us a
completely wrong region for $\dcp$.
Thus the determination of the mass hierarchy is important
for the measurement of $\dcp$.
The third one is the octant degeneracy\,\cite{Fogli:1996pv}.
The main contribution to the probabilities
$1-P(\nu_\mu\to\nu_\mu)$ and $1-P(\bar{\nu}_\mu\to\bar{\nu}_\mu)$
is proportional to $\sin^22\theta_{23}$, and
we can only determine $\sin^22\theta_{23}$ from
the disappearance channel.
If $\theta_{23}$ is not maximal, then
we have two possibilities
$\cos2\theta_{23}>0$ and
$\cos2\theta_{23}<0$, so this ambiguity gives us
two solutions $(\theta_{23}, \dcp)$ and
$(90^\circ-\theta_{23}, \dcp^\prime)$.
The octant degeneracy can create a source
of the uncertainty in $\theta_{23}$, so
its resolution is important for precise
measurement of $\theta_{23}$. 
These three types of degeneracies together created a eight fold degeneracy degeneracy \cite{Barger:2001yr} when the precise value of $\theta_{13}$ was unknown. After the discovery of $\theta_{13}$, now the 
eight fold degeneracy breaks into a four fold degeneracy. 
At present the relevant degeneracy in the neutrino oscillation probability is known as the ``generalized hierarchy-octant-$\dcp$ degeneracy" \cite{Ghosh:2015ena}. This generalized degeneracy 
consists of hierarchy-$\dcp$ degeneracy \cite{Prakash:2012az} and octant-$\dcp$ degeneracy \cite{Agarwalla:2013ju} which we discuss in the next section for the T2HK and DUNE baselines.

\subsection{T2HK and DUNE}
\label{t2hk-dune}

T2HK is the long baseline
experiment which is planned in Japan, and its baseline length and peak
energy is $L$=295 km, $E\sim 0.6$ GeV, respectively.  On the other hand,
DUNE is another long baseline experiment which is planned in USA,
and its baseline length and peak
energy is $L$=1300 km, $E\sim3$ GeV, respectively.
The matter effect appears in the neutrino oscillation probability
typically in the form of
$G_FN_eL/\sqrt{2}$ = $[\rho/(2.6g/cm^3)][L/(4000\mbox{\rm km})]$.
The baseline length of T2HK is too short for the matter
effect 
so T2HK has poor sensitivity to the mass hierarchy. Hence
the sign degeneracy can be in principle serious in
determination of $\dcp$ at T2HK.
On the other hand, the baseline length of DUNE is
comparable to the typical length which is estimated
by the matter effect, so DUNE is expected to be
sensitive to the mass hierarchy.

As was discussed in the case of T2K and \nova in Ref. \cite{Prakash:2012az},
we can see whether this so called hierarchy-$\dcp$ degeneracy can be resolved or not
by looking at the behaviors of the appearance channel probabilities.
In Fig.\,\ref{t2hk-dune-hierarchy}, we plot the appearance channel probability spectrum for a fixed value of $\theta_{23}=45^\circ$. 
The first and second panels are for neutrino and antineutrino probabilities of T2HK and the third and fourth panels are for neutrino and antineutrino probabilities of DUNE respectively. 
In each panel, the red band correspond to NH and the blue band
corresponds to IH. The width of the band is due to the variation of the phase $\dcp$.
The overlap region between the red band and the blue band correspond to the hierarchy-$\dcp$ degeneracy.
From the Fig.\,\ref{t2hk-dune-hierarchy} we see that
that in the case of $\dcp=-90^\circ$ with
the normal mass hierarchy (NH) or $\dcp=90^\circ$ with
the inverted mass hierarchy (IH), T2HK can
resolve the mass hierarchy by the neutrino mode $\nu_\mu\to\nu_e$
alone. These are the favorable values of $\dcp$ where there is no hierarchy-$\dcp$ degeneracy. On the other hand, in the case of $\dcp=90^\circ$ with
the normal mass hierarchy or $\dcp=-90^\circ$ with
the inverted mass hierarchy,
we see that T2HK cannot
resolve the mass hierarchy even if we combine the neutrino mode $\nu_\mu\to\nu_e$
and the antineutrino mode $\bar{\nu}_\mu\to\bar{\nu}_e$, because
the curve for $\dcp=90^\circ$ with NH or
$\dcp=-90^\circ$ with IH lies in the middle of the overlapping region
in the both modes. These are the unfavorable values of $\dcp$ which suffers from the hierarchy-$\dcp$ degeneracy.
For the case of DUNE, though the behavior of the appearance channel probabilities are the same as that of T2HK, but the NH and IH bands are quite well separated. 
For this reason it is possible to have hierarchy sensitivity in DUNE even for the unfavorable parameter values of $\dcp$.

On the other hand, the situation of the
octant degeneracy is different.
As shown in \cite{Agarwalla:2013ju}, one can identify the octant-$\dcp$ degenerate parameter space by looking at the appearance channel probabilities.
In Fig.\,\ref{t2hk-dune-octant} we plot the same as that of Fig.\,\ref{t2hk-dune-hierarchy} but for a fixed value of $\Delta m_{31}^2$. We assume the normal mass hierarchy
for simplicity. In these plots the red band corresponds to LO ($\theta_{23}=42^\circ$) and the blue 
band correspond to HO ($\theta_{23}=48^\circ$). The width of the bands are due to the variation of $\dcp$.
From Fig.\,\ref{t2hk-dune-octant} we see that,
in the case of $\dcp=90^\circ$, if $\theta_{23}$ lies in
the first octant (i.e., if $\theta_{23}=42^\circ$), then both
T2HK and DUNE can resolve the octant degeneracy by
the neutrino mode $\nu_\mu\to\nu_e$ alone, and
if $\theta_{23}$ lies in
the second octant (i.e., if $\theta_{23}=48^\circ$), then
T2HK and DUNE can resolve the octant degeneracy by
the antineutrino mode $\bar{\nu}_\mu\to\bar{\nu}_e$ alone.
This argument also applies to the case of $\dcp=-90^\circ$,
and we conclude that both T2HK and DUNE has potential
to resolve the octant degeneracy by themselves by
combining the neutrino and antineutrino modes. These conclusions also hold for the inverted hierarchy
\footnote{ Note that our discussion of degeneracy is valid only for a fixed value of energy $E$. In the realistic 
scenario the nature of these degeneracies can be different due to the energy dependence of the probability spectrum.}.

\subsection{Atmospheric neutrinos \label{atm}}

\begin{figure*}
\begin{center}
\begin{tabular}{lr}
\hspace{-0.1in}
\includegraphics[width=0.25\textwidth]{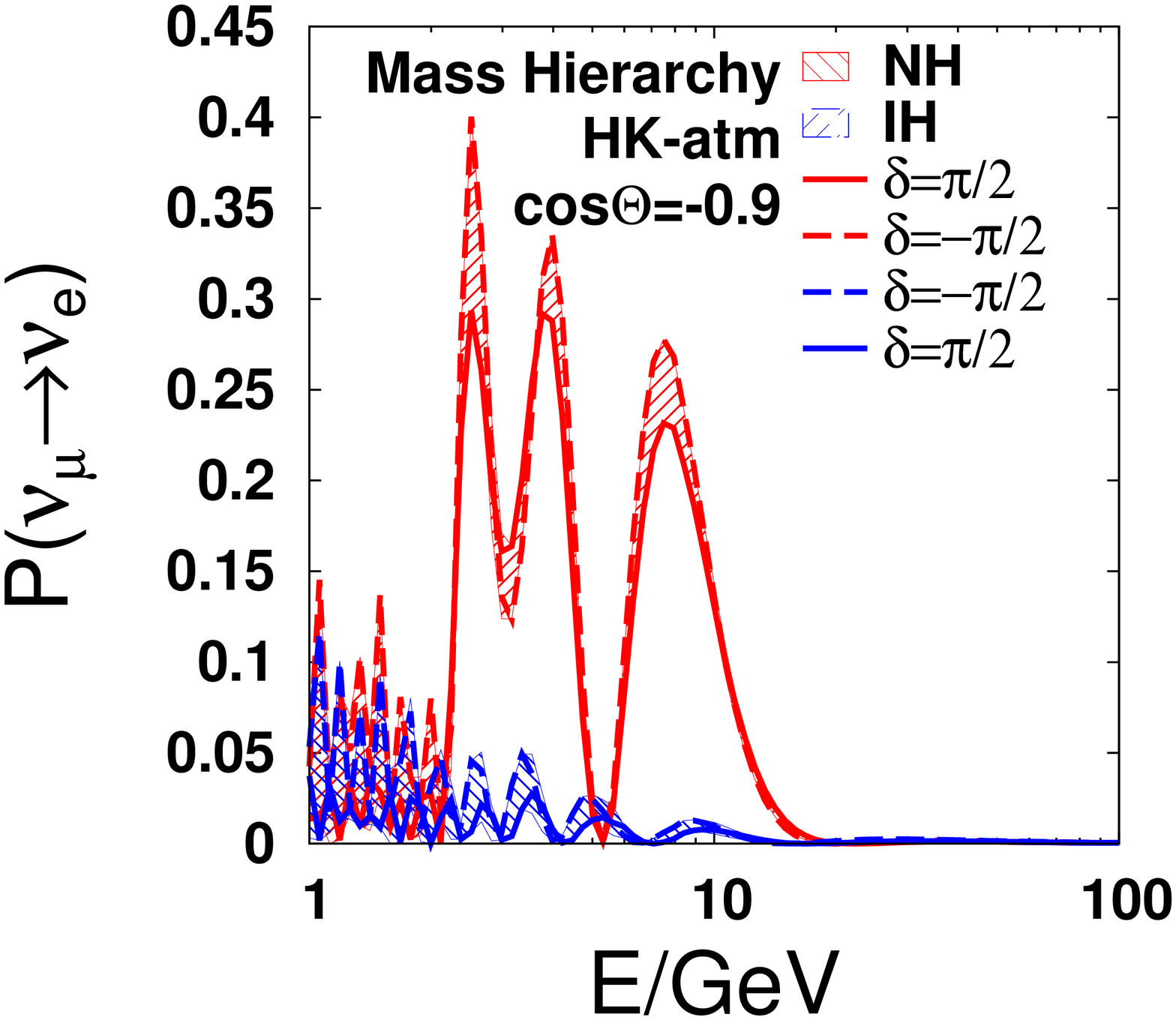}
\hspace*{-0.1in}
\includegraphics[width=0.25\textwidth]{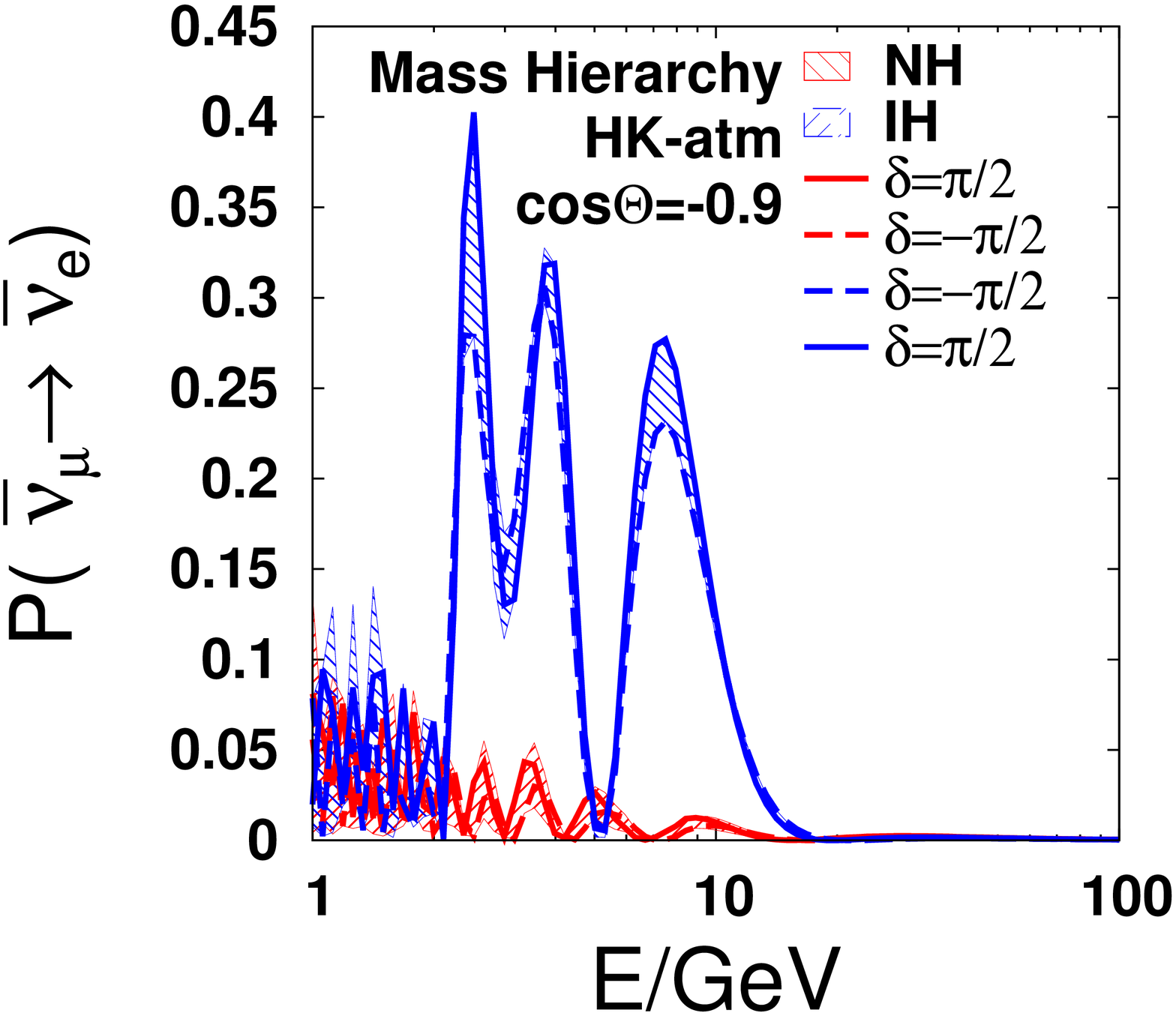}
\hspace{-0.1in}
\includegraphics[width=0.25\textwidth]{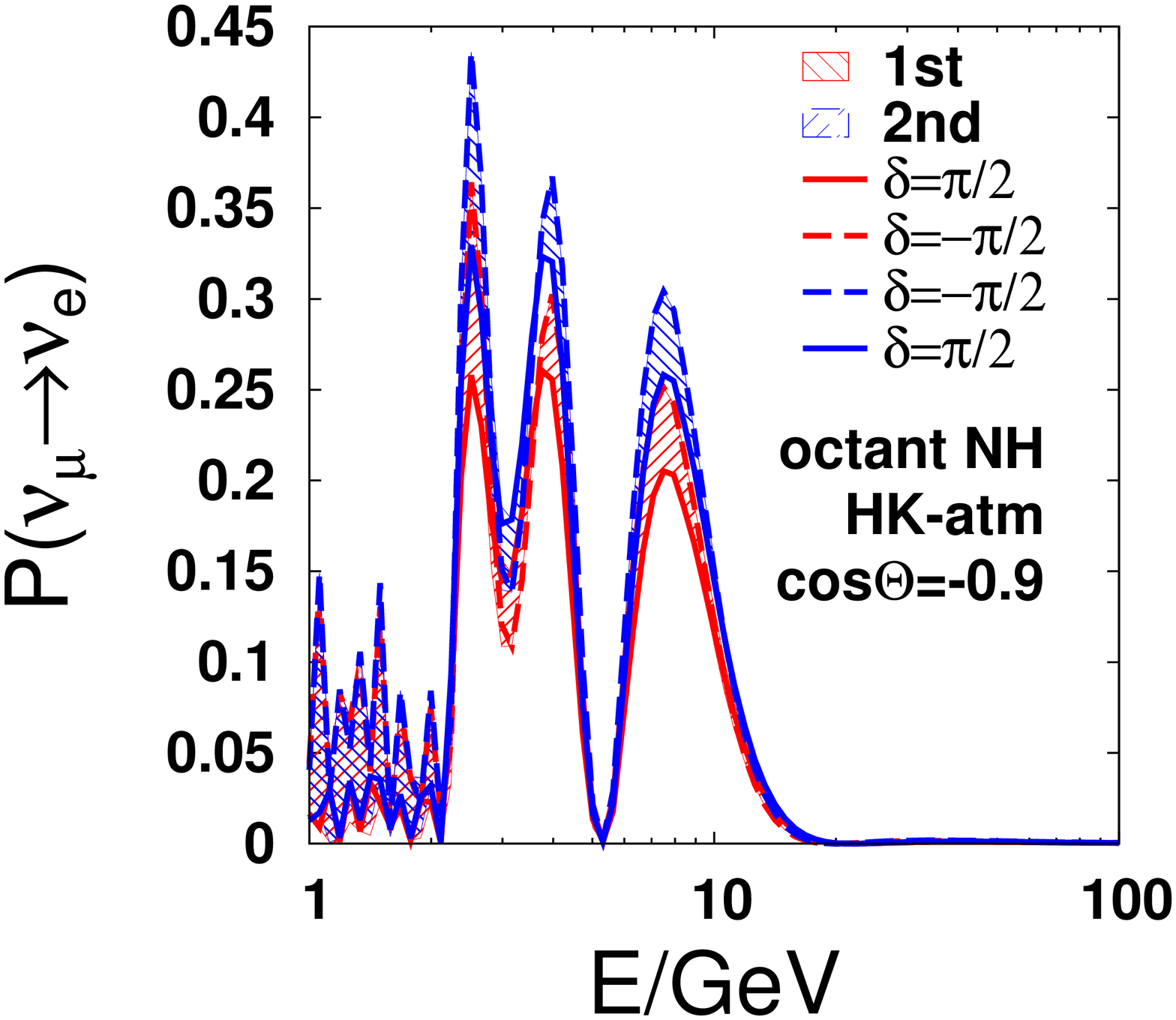}
\hspace{-0.1in}
\includegraphics[width=0.25\textwidth]{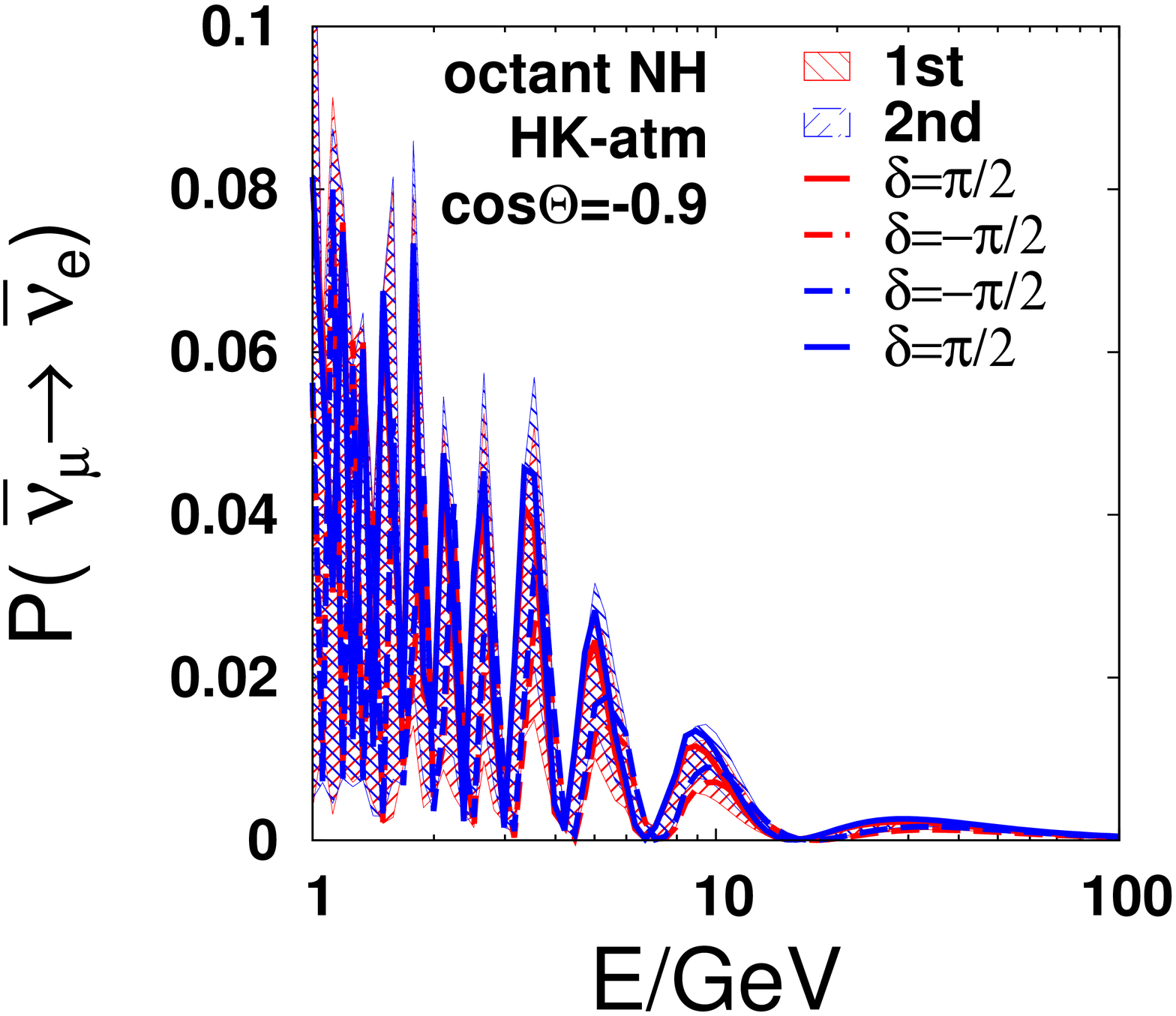}
\end{tabular}
\end{center}        
\caption{The behaviors of the appearance probabilities
$P(\nu_\mu\to\nu_e)$ (first and third panels) and
$P(\bar{\nu}_\mu\to\bar{\nu}_e)$ (second and fourth panels)
for the atmospheric neutrino measurements for
the zenith angle $\cos\Theta=-0.9$
for the two different mass hierarchies (left two panels)
and in the
two different octants (right two panels).
In the right two panels $\theta_{23}=42^\circ$ or
$\theta_{23}=48^\circ$ is taken as a reference value, and
Normal hierarchy is assumed.
}
\label{hk-atm-prob}
\end{figure*}

The measurement of atmospheric neutrinos at Hyperkamiokande (HK)
gives us information which is
complimentary to long-baseline experiments.
Its advantage is that we can get the information
on the oscillation probabilities for wide ranges
of the neutrino energy and the baseline length,
particularly for more than 4000 km,
so the measurement gives us information on
the matter effect.  On the other hand,
its disadvantages are that there are uncertainties
on the baseline length, that it is difficult to
distinguish neutrinos and antineutrinos,
and that the oscillation
probability has to be deduced indirectly from 
the measured flux, because
the measured flux
$F(\nu_\alpha)=F_0(\nu_e)P(\nu_e\to\nu_\alpha)+
F_0(\nu_\mu)P(\nu_\mu\to\nu_\alpha)$ $~(\alpha=e, \mu)$
is the sum of the each flux
which are produced out of the original flux
$F_0(\nu_\alpha)~(\alpha=e, \mu)$ through neutrino oscillations.

In our simulation of the atmospheric neutrino
measurements at Hyperkamiokande, the data
which is most sensitive to the mass hierarchy
and the octant degeneracy is the e-like
multi-GeV events\,\footnote{In the multi-GeV energy
region ($\sim$ 10\,GeV), the ratio of the original
e-like flux $F_0(\nu_e)$ to the original $\mu$-like one
$F_0(\nu_\mu)$ is approximately
1:6, so the relative contribution of the
appearance probability in
the total flux after oscillations is
larger in the e-like flux
$F(\nu_e)\sim F_0(\nu_e)
\{P(\nu_e\to\nu_e)+6P(\nu_\mu\to\nu_e)\}$
than in the $\mu$-like one
$F(\nu_\mu)\sim F_0(\nu_\mu)
\{(1/6)P(\nu_e\to\nu_\mu)+P(\nu_\mu\to\nu_\mu)\}$.
Therefore the e-like events are more sensitive
to the mass hierarchy than the $\mu$-like ones.} 
in the higher energy region
for the zenith angle $-1.0 < \cos\Theta < -0.8$.
The disappearance probability
$P(\nu_e\to\nu_e)$ is insensitive to
the matter effect,
so what makes the difference
between the two mass hierarchies is the
appearance probability $P(\nu_\mu\to\nu_e)$.
In Fig.\,\ref{hk-atm-prob}, the appearance
probabilities are plotted for $\cos\Theta = -0.9$. The first two panels are similar as that of Fig.\,\ref{t2hk-dune-hierarchy}
and the third and fourth panels are similar as that of Fig.\,\ref{t2hk-dune-octant}.  
From the first two panels, it is clear that
as far as the appearance probabilities
are concerned, we should be able to determine the mass hierarchy
on the condition that we can measure the
appearance probabilities for the energy and the
baseline length in the atmospheric neutrino
measurements.
As for the octant, third panel and fourth panels
of Fig.\,\ref{hk-atm-prob} suggests that
the atmospheric neutrino measurements may
resolve the octant degeneracy as it combines both neutrino and antineutrino probabilities.

\section{Analysis}
\label{analysis}

\subsection{Simulation Details}
Here we assume the following parameters
for each experiment:

\begin{enumerate}
\renewcommand{\labelenumi}{\arabic{enumi})}
\item T2HK: We have taken the
parameters from Ref.\,\cite{Abe:2014oxa}.\footnote{
In the recent design of the Hyperkamiokande
project\,\cite{kajita:2016}, two tanks
with the fiducial volume 0.19 Mton each
with the staging construction is planned.
At this moment, 
the details are not known, so we will
analyze T2HK and the atmospheric neutrino
measurements at Hyperkamiokande
with the parameters in the
old design of Hyperkamiokande
throughout this paper.
}

Our $\chi^2$ is a function of
the two sets of the oscillation parameters
$\vec{p}_{\mbox{\rm ex}}$ and $\vec{p}_{\mbox{\rm th}}$, and is
defined as:
 \begin{eqnarray*}
 \chi^2_{\mbox{\rm\scriptsize \ T2HK}} (\vec{p}_{\mbox{\rm ex}},\vec{p}_{\mbox{\rm th}}) 
=  \min_{\{\xi_j\}}  \bigg[ \chi^2_{\mbox{\rm\scriptsize \ T2HK}} (\vec{p}_{\mbox{\rm ex}},\vec{p}_{\mbox{\rm th}}; \{\xi_j\}) 
 + {\displaystyle \sum_j} \left(\frac{\xi_j}{\pi^j}\right)^2 \bigg]\,,
\end{eqnarray*}
where $\{ \xi_j \}$ are the pull variables, 
$\pi^j$ are the ($1\sigma$) systematic errors for the pull variable $\xi_j$, with
\begin{eqnarray}
  \chi^{2 \ }_{\mbox{\rm\scriptsize \ T2HK}} (\vec{p}_{\mbox{\rm ex}},\vec{p}_{\mbox{\rm th}}; \{\xi_j\})
= 2 {\displaystyle \sum_i} \Bigg[ M_{i}(\vec{p}_{\mbox{\rm th}}) - N_{i}(\vec{p}_{\mbox{\rm ex}}) 
+N_{i}(\vec{p}_{\mbox{\rm ex}}) \ln \left( {\displaystyle \frac{N_{i}(\vec{p}_{\mbox{\rm ex}})}
{M_{i}(\vec{p}_{\mbox{\rm th}})} } \right) \Bigg] ~. 
\label{chi2t2hk}
 \end{eqnarray}
 The `experimental' data $N_{i}(\vec{p}_{\mbox{\rm ex}})$ are simulated using the `true' oscillation parameters 
 $\vec{p}_{\mbox{\rm ex}}$, while the `theoretical' events $N_{i}(\vec{p}_{\mbox{\rm th}})$ are generated 
 using the `test' oscillation parameters $\vec{p}_{\mbox{\rm th}}$. The subscript $i$ here 
 runs over all the energy bins. The theoretical events get modified due to systematic errors as
\begin{eqnarray*}
 M_{i} (\vec{p}_{\mbox{\rm th}})
 =  N_{i} (\vec{p}_{\mbox{\rm th}}) \Bigg[ 1 + {\displaystyle \sum_k} \xi_k
 + {\displaystyle \sum_l}\xi_l {\displaystyle \frac{E_i-E_{\mbox{\rm av}}}{E_{\mbox{\rm max}}-E_{\mbox{\rm min}}} } \Bigg] \,.
 \end{eqnarray*}
are the theoretical numbers of events
including the systematic uncertainty.
Where the index $k(l)$ runs over the relevant normalization (tilt) 
 systematic errors for a given 
 experimental observable. 
The normalization errors affect the scaling of events and 
the tilt errors affect the energy dependence of the events. 
 All the pull variables $\{ \xi_j \}$ take values in the range 
 $(-3\pi^j,3\pi^j)$, 
so that the errors can vary from $-3\sigma$ to $+3\sigma$. Here, $E_i$ 
 is the mean energy of the $i^{\rm th}$ energy bin, $E_{\mbox{\rm min}}$ and $E_{\mbox{\rm max}}$ are the 
 limits of the full energy range, and $E_{\mbox{\rm av}}$ is their average. 
The final $\chi^2$ is then calculated 
by minimizing over all combinations of $\xi_j$. 



Total exposure: 1.56 $\times 10^{22}$ POT (protons on target). This implies, for a beam power of $1 \times 10^{21}$ POT/year, it corresponds to 15.6 year running. 

Flux: 30 GeV proton beam

Detector: 0.56 Mton (fiducial volume) water \cnv\

$\nu$:$\bar{\nu}$ = 1:1

Systematics: an overall normalization error of 3.3\% for both appearance and disappearance channel in neutrino mode and 
6.2\% (4.5\%) for appearance (disappearance) channel in antineutrino mode. The normalization error is same for both signal and background.
For tilt error we have taken 1\% (5\%) corresponding to signal (background)
in appearance channel and 0.1\% in disappearance channel for both signal and background. Tilt error for neutrinos and antineutrinos are taken to be same.

For T2HK we have reproduced the sensitivity as given in Ref.\,\cite{Abe:2014oxa}.  

\item DUNE: We have taken the
parameters from Ref.\,\cite{Acciarri:2015uup}.

 In the case of the DUNE experiment
$\chi^2$ is defined as in the same manner as T2HK.

Total exposure: 1.0 $\times 10^{22}$ POT. Thus for a beam power of $1 \times 10^{21}$ POT/year, it will run for 10 years.

Flux: 120 GeV proton beam

Detector: 34 kton (fiducial volume) Liquid Argon

$\nu$:$\bar{\nu}$ = 1:1

Systematics: an overall normalization error of 2\% (10\%) for appearance channel and 5\% (15\%) for disappearance channel corresponding to signal (background) in both neutrino and 
antineutrino mode. The tilt error is 2.5\%. 

Our sensitivity results of DUNE are in agreement with Ref.\,\cite{Acciarri:2015uup}.

The simulations of T2HK and DUNE
have been performed with
the software GLoBES \cite{Huber:2004ka,Huber:2007ji}.

\item Atmospheric neutrino measurements at Hyperkamiokande:
We have taken the parameters from
Ref.\,\cite{hkloi}.

Detector: 0.56 Mton (fiducial volume) water \cnv\

Duration: 200 days $\times$ 10 years = 2000 days

As in Ref.\,\cite{Fukasawa:2015jaa},
we use the data set of the sub-GeV events with the two energy-bins,
the multi-GeV events with the two energy-bins, and
the combined stopping and through-going upward going $\mu$ events
with the single energy-bin,
where the number of the zenith angle bins is ten for all these
channels.

Simulation of the atmospheric neutrino
at Hyperkamiokande is done with
the codes which were
used in Refs.\,\cite{Foot:1998iw,Yasuda:1998mh,Yasuda:2000de,Fukasawa:2015jaa}.
In our analysis, we compared our $\chi^2$
and the value of $\chi^2$ given by the HK
collaboration in Ref.\,\cite{Abe:2014oxa}
and normalized our $\chi^2$
for each analysis (mass hierarchy, octant,
and CP violation) so that the
two $\chi^2$ approximately coincide with each other.
Thus our sensitivity of the HK experiment matches with  Ref.\,\cite{Abe:2014oxa}.

The analysis was performed using $\chi^2$-method.
$\chi^2$ depends on
the two sets of the oscillation parameters
$\vec{p}_{\mbox{\rm ex}}$ and $\vec{p}_{\mbox{\rm th}}$, and is
defined as
\begin{eqnarray}
&{\ }&\hspace*{-5mm}
\chi^2_{\rm atm}(\vec{p}_{\mbox{\rm ex}},\vec{p}_{\mbox{\rm th}})
\nonumber\\
&{\ }&\hspace*{-10mm}
 =
\min_{\theta_{23},|\Delta m^2_{32}|,\delta,\epsilon_{\alpha\beta}}
\left[
\chi_{\rm sub-GeV}^2(\vec{p}_{\mbox{\rm ex}},\vec{p}_{\mbox{\rm th}}) 
+\chi_{\rm multi-GeV}^2(\vec{p}_{\mbox{\rm ex}},\vec{p}_{\mbox{\rm th}}) 
+\chi_{\rm upward}^2(\vec{p}_{\mbox{\rm ex}},\vec{p}_{\mbox{\rm th}}) 
\right]\,,
\label{eqn:chi}
\end{eqnarray}
where
$\displaystyle
\chi_{\rm sub-GeV}^2(\vec{p}_{\mbox{\rm ex}},\vec{p}_{\mbox{\rm th}})$,
$\displaystyle
\chi_{\rm multi-GeV}^2(\vec{p}_{\mbox{\rm ex}},\vec{p}_{\mbox{\rm th}})$
and $\displaystyle
\chi_{\rm upward}^2(\vec{p}_{\mbox{\rm ex}},\vec{p}_{\mbox{\rm th}})$
are $\chi^2$ for the sub-GeV, muti-GeV, and upward $\mu$
events, and they are the same as $\chi_{\rm sub-GeV}^2$,
$\chi_{\rm multi-GeV}^2$ and $\chi_{\rm upward}^2$
defined in Ref.\,\cite{Fukasawa:2015jaa}, respectively.
The way we introduce the systematic errors is an extension of that
in the original analysis of atmospheric neutrinos.\,\cite{Fukuda:1998mi}
In our code we have used the old atmospheric neutrino
flux\,\cite{Honda:1995hz} instead of the new
one\,\cite{Honda:2015fha}, because our simulation results become
closer to the ones by Superkamiokande collaboration with the old flux.
However the difference in $\chi^2$ between the old flux and
the new flux is small ($\sim$ a few \%),
since our simulated experimental data are also evaluated with
the same flux.  So the choice of the atmospheric neutrino flux does not affect
our conclusions very much.
We have set the systematic errors to the same values as in Ref.\,\cite{Ashie:2005ik} except several unimportant factors.  
In particular, we confirmed that taking a uncertainty in the $E_{\nu}$ spectral index which is omitted 
in our analysis into consideration gives negligible contributions to $\chi^2$.
See Refs.\,\cite{Fukasawa:2015jaa,Fukasawa:2016nwn} for further details.

\end{enumerate}

In our simulation we have fixed the value of $\sin^2\theta_{12}=0.3$, $\sin^2 2 \theta_{13}=0.1$, $\Delta m_{21}^2 =  7.5 \times 10^{-3}$ eV$^2$ 
and $|\Delta m_{{\rm eff}}^2| = 2.4 \times 10^{-3}$ eV$^2$\footnote{
Following Ref.\,\cite{Nunokawa:2005nx}, we adopt the
effective mass squared difference for the
$\mu$ disappearance channel, and it is defined as
$\Delta m^2_{\mbox{\scriptsize\rm eff}}\equiv s^2_{12}\Delta m^2_{31}
+c^2_{12}\Delta m^2_{32}+\cos\dcp\, 
s_{13}\sin2\theta_{12}\tan\theta_{23}\,\Delta m^2_{21}$. We consider this in our analysis because in vacuum, the hierarchy degeneracy do not occur for 
$P_{\mu \mu}(\Delta m^2_{31}) = P_{\mu \mu}(-\Delta m^2_{31})$ but it occurs for $P_{\mu \mu}(\Delta m^2_{{\rm eff}}) = P_{\mu \mu}(-\Delta m^2_{{\rm eff}})$.}. 
This is a good approximation since these parameters are well constrained from the global analysis from the world neutrino data \cite{Capozzi:2013csa,Forero:2014bxa,Bergstrom:2015rba} and 
thus marginalization over these parameters will not affect
the sensitivity much.
We vary $\theta_{23}$ from $40^\circ$ to $50^\circ$ and $\dcp$ from $-180^\circ$ to $+180^\circ$. For estimating the precision of $\theta_{23}$ and $\Delta m^2_{{\rm eff}}$, 
we have kept $\dcp$ fixed
at $-90^\circ$ for both true and test spectrum.

\begin{figure*}
\begin{tabular}{lr}
\hspace{-0.6 in}
\includegraphics[width=0.6\textwidth]{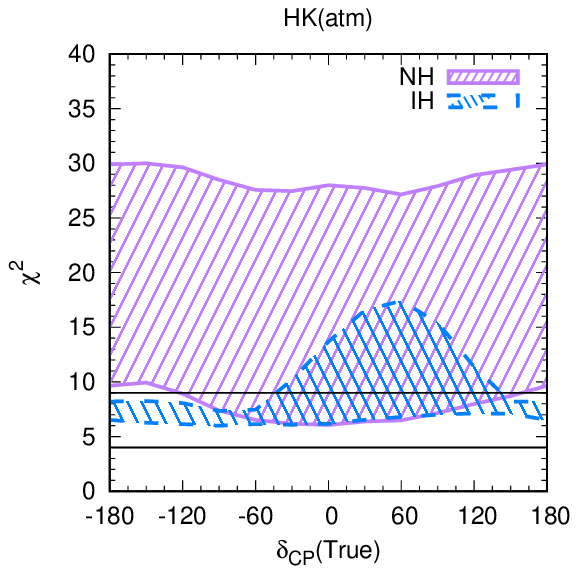} 
\hspace{-1.4 in}
\includegraphics[width=0.6\textwidth]{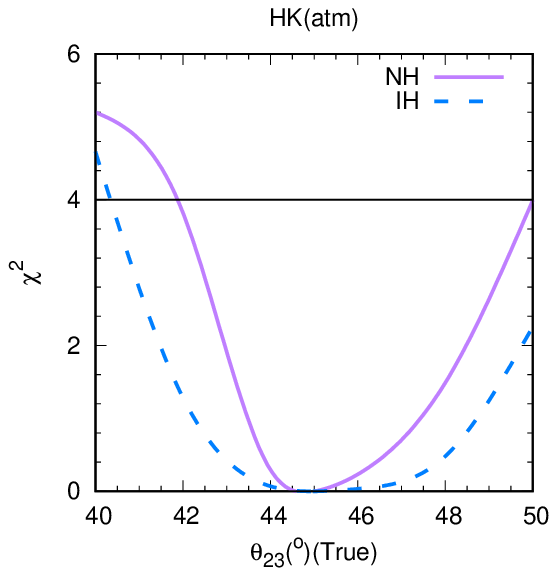} \\ 
\hspace{-0.6 in}
\includegraphics[width=0.6\textwidth]{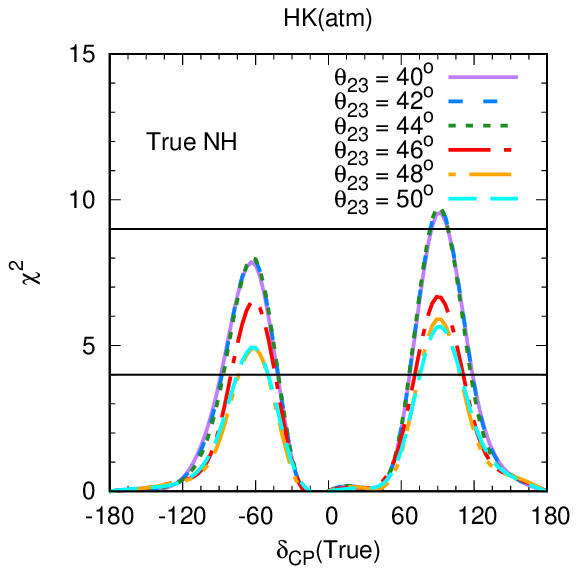}
\hspace{-1.4 in}
\includegraphics[width=0.6\textwidth]{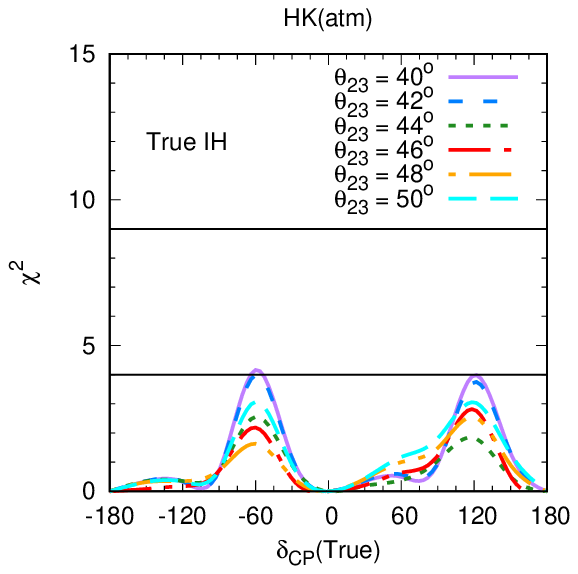}
\end{tabular}
\caption{Sensitivity of the HK atmospheric neutrino experiment
to the mass hierarchy (the top left panel), octant (the top right panel)
and CP violation (bottom row).
In the top left panel the width of the band comes from the
uncertainty in $\theta_{23}$.
}
\label{hk-atm-chi2}
\end{figure*}

\subsection{Sensitivity of HK}

\begin{figure*}
\begin{center}
\begin{tabular}{lr}
\hspace{-0.6 in}
\includegraphics[width=0.6\textwidth]{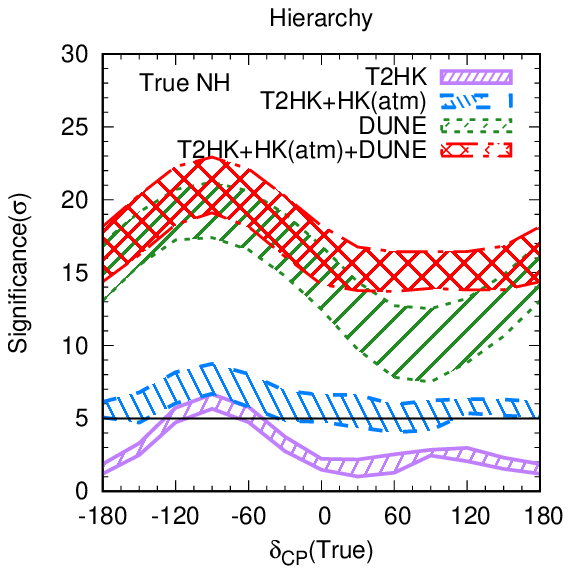} 
\hspace*{-1.4in}
\includegraphics[width=0.6\textwidth]{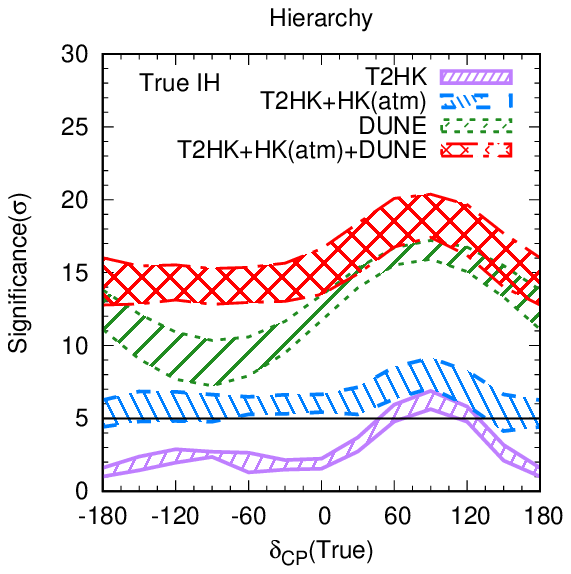} \\
\hspace{-0.6 in}
\includegraphics[width=0.6\textwidth]{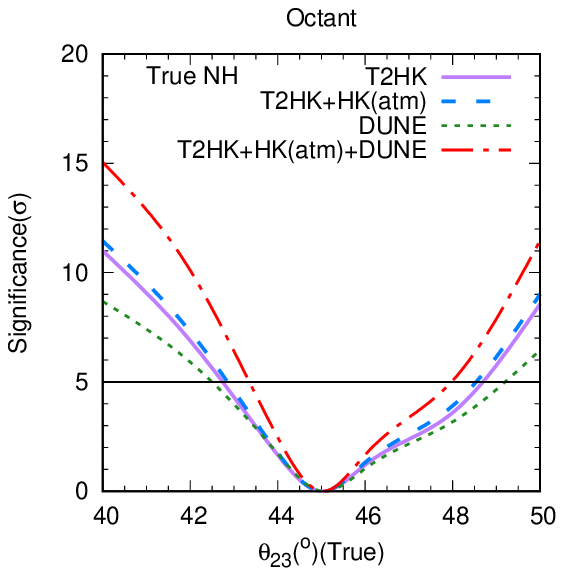} 
\hspace{-1.4 in}
\includegraphics[width=0.6\textwidth]{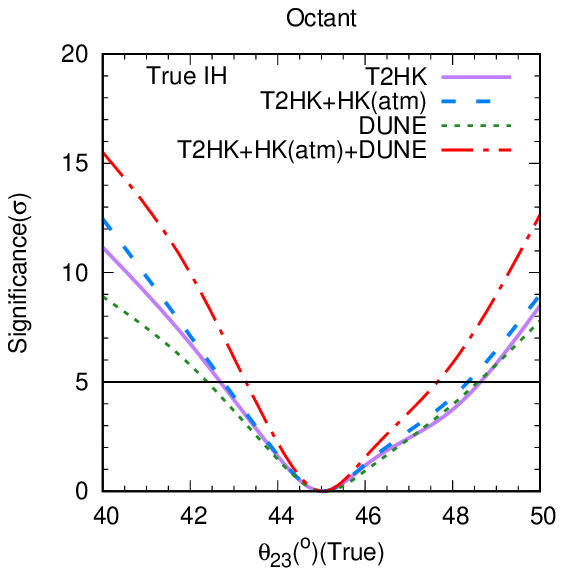} \\
\hspace{-0.6 in}
\includegraphics[width=0.6\textwidth]{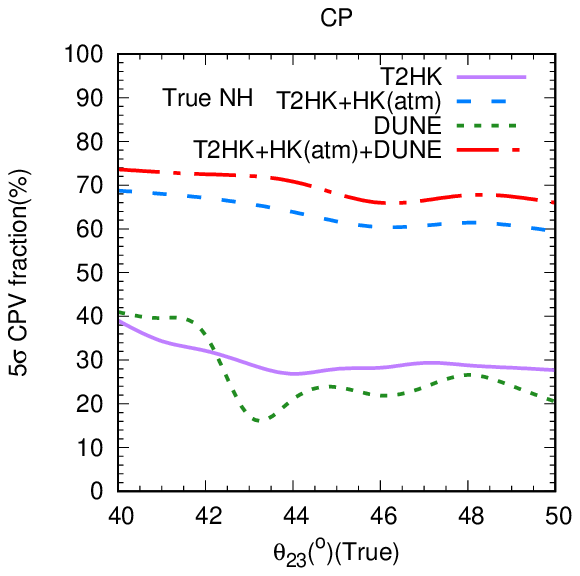} 
\hspace{-1.4 in}
\includegraphics[width=0.6\textwidth]{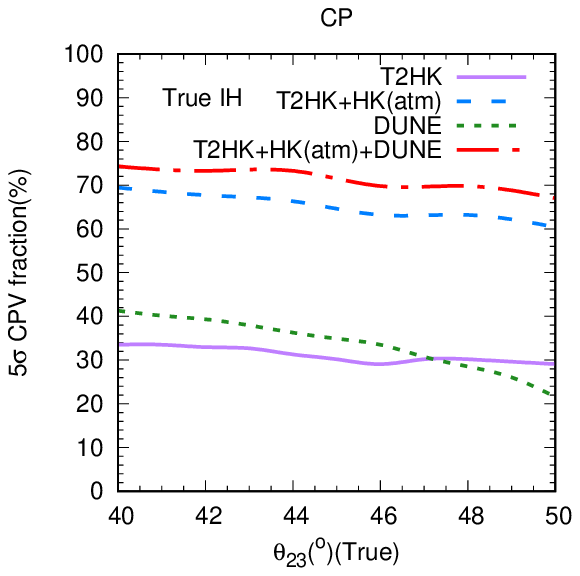}
\end{tabular}
\end{center}
\caption{Sensitivity to the mass hierarchy, octant and CP violation.
In the upper and middle rows the vertical axis is $\sqrt{\chi^2}$.
}
\label{hierarchy}
\end{figure*}

\begin{figure*}
\begin{tabular}{lr}
\hspace{-0.6 in}
\includegraphics[width=0.6\textwidth]{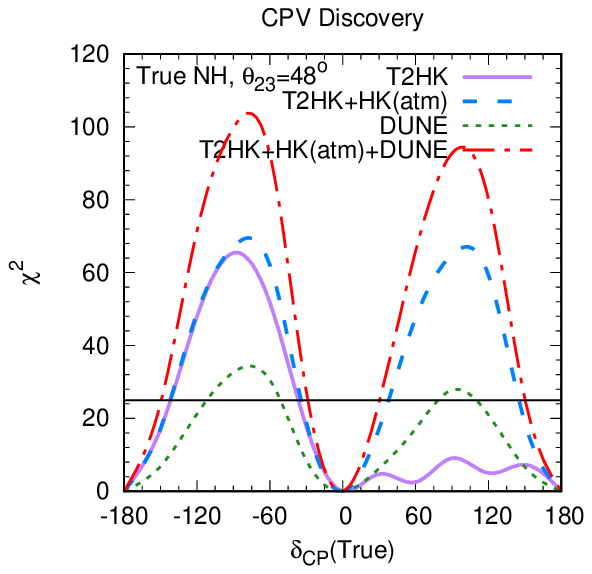}
\hspace*{-1.4in}
\includegraphics[width=0.6\textwidth]{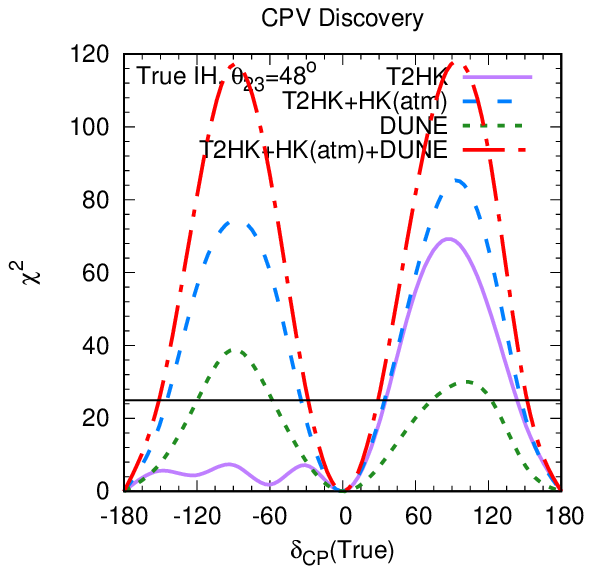} \\
\end{tabular}
\caption{Sensitivity to CP violation for $\theta_{23}=48^\circ$.
}
\label{cp}
\end{figure*}

In this section we study the potential of the HK experiment to determine the neutrino mass hierarchy, octant and $\dcp$. 

\begin{figure*}
\begin{tabular}{lr}
\vspace{-0.2 in}
\includegraphics[width=0.5\textwidth]{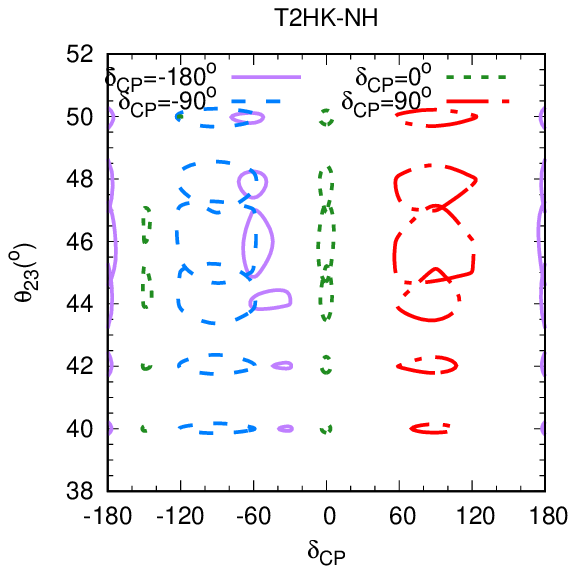}
\hspace*{-1.0in}
\includegraphics[width=0.5\textwidth]{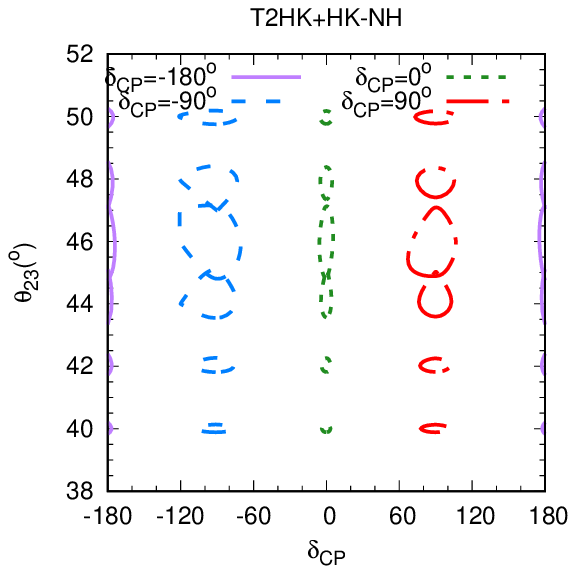} \\
\vspace{-0.2 in}
\includegraphics[width=0.5\textwidth]{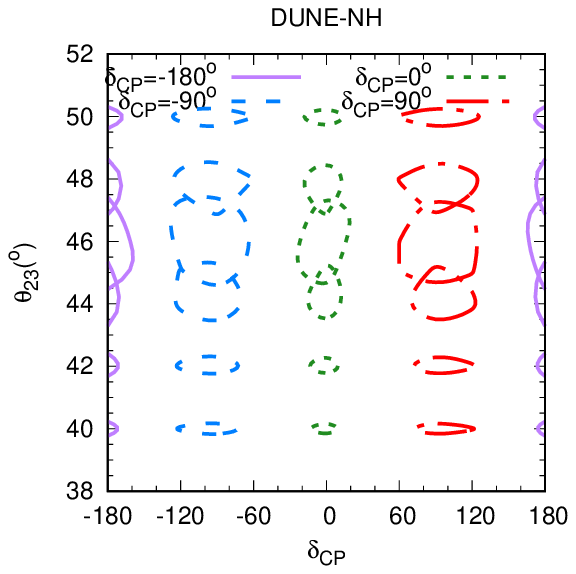}
\hspace*{-1.0in}
\includegraphics[width=0.5\textwidth]{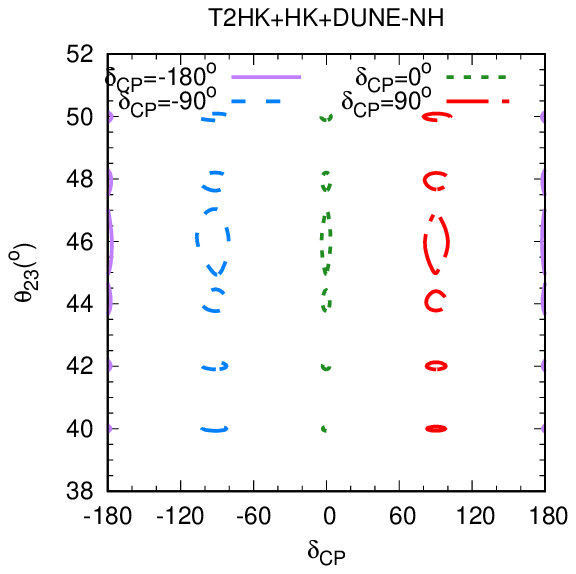} \\ 
\vspace{-0.2 in}
\includegraphics[width=0.5\textwidth]{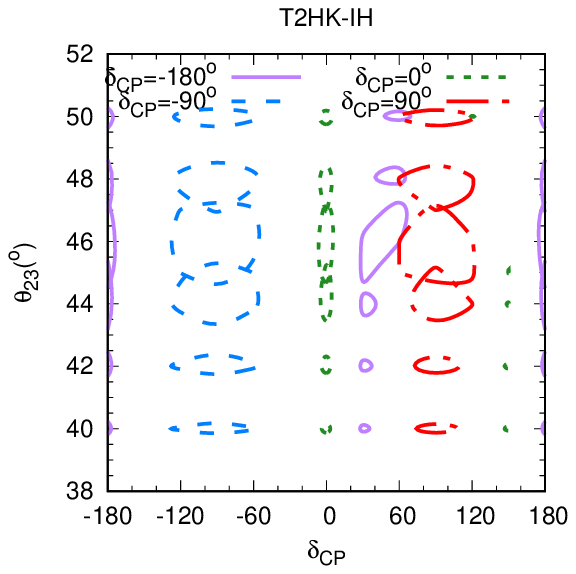}
\hspace*{-1.0in}
\includegraphics[width=0.5\textwidth]{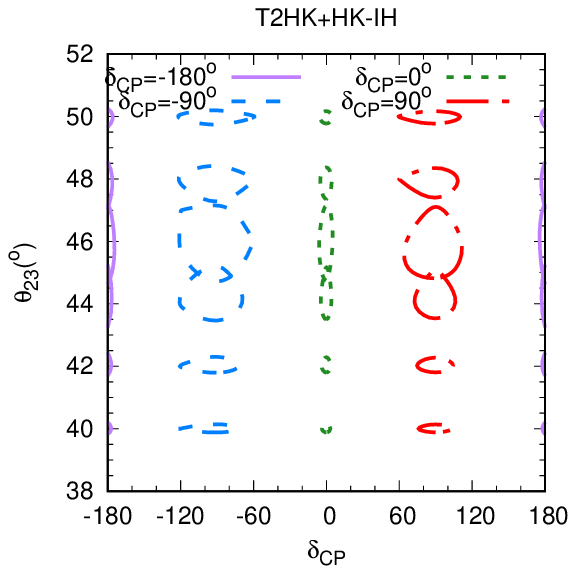} \\ 
\includegraphics[width=0.5\textwidth]{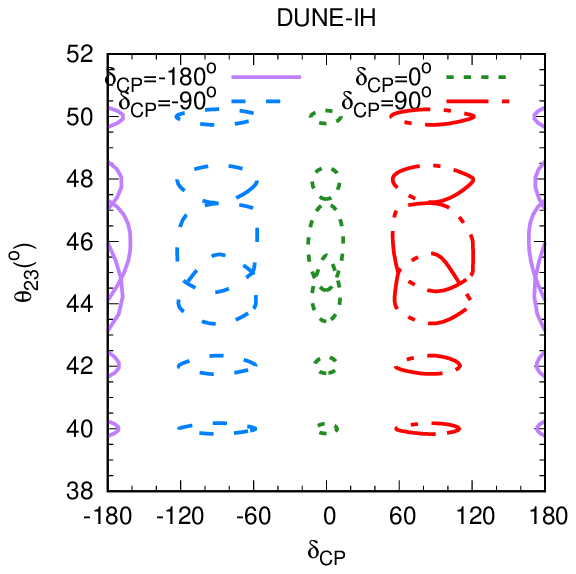}
\hspace*{-1.0in}
\includegraphics[width=0.5\textwidth]{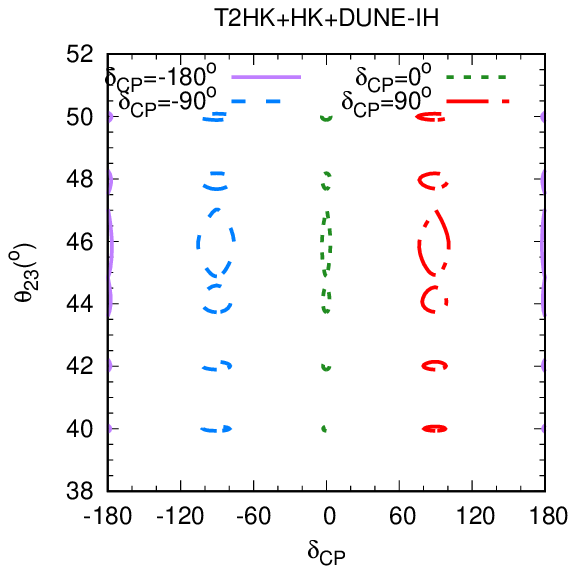} 
\end{tabular}
\caption{90\% C.L. contour in the test
$\dcp$ and test
$\theta_{23}$ plane for several representative true points.}
\label{param_precision}
\end{figure*}


Fig.\,\ref{hk-atm-chi2} shows the
sensitivity of HK to the mass hierarchy,
octant and CP violation.
In the top left panel we present the hierarchy sensitivity $\chi^2$ as a function of true $\dcp$ for the HK experiment. Hierarchy sensitivity of an experiment is defined by
taking the right hierarchy in the true spectrum and wrong hierarchy in the test spectrum. While calculating $\chi^2$ we have marginalized over $\theta_{23}$ and $\dcp$ in the test spectrum.
The purple band in the figure corresponds to sensitivity in NH and the blue band corresponds to
sensitivity in IH. The width of the band is due to the variation of true $\theta_{23}$ from $40^\circ$ to $50^\circ$ which is the current allowed values of 
$\theta_{23}$. From the figure we observe that, for conservative value of $\theta_{23}$,\footnote{The word ``conservative (optimistic)" means the value of $\theta_{23}$ for which the 
hierarchy sensitivity is minimum (maximum).}
the hierarchy $\chi^2$ is close to 6 around $\dcp = 0^\circ$ for both NH and IH and for optimistic value of $\theta_{23}$, 
the hierarchy sensitivity increases to $\chi^2=30$ for NH around $\pm 180^\circ$
and $\chi^2 = 18$ for IH around $\dcp=+60^\circ$. Here it is interesting to see that the hierarchy sensitivity of IH is 
in general poorer as compared to NH. It is also important to note that
for $-180^\circ < \dcp < -60^\circ$, the width of the IH band is very narrow. This is because for water \cnv\ detectors, 
it has been shown that NH-LO is degenerate with IH-HO \cite{Winter:2013ema,Capozzi:2015bxa}.
Thus because of the octant degeneracy, we obtain a very poor hierarchy sensitivity for HK in IH in the above mentioned parameter space. Note that although NH is not free from this degeneracy 
but the degeneracy in NH seems less severe as compared to the degeneracy in IH. We will discuss this point in detail in the appendix.

In the top right panel of Fig.\,\ref{hk-atm-chi2} we plot the octant $\chi^2$ as a function of the true $\theta_{23}$. The sensitivity to the octant
is defined by taking the correct octant of $\theta_{23}$ in the true spectrum and wrong octant of $\theta_{23}$ in the test spectrum. In the process of calculating $\chi^2$ 
for octant sensitivity we have marginalized over sign($\Delta m^2_{\mbox{\scriptsize\rm eff}}$) and $\dcp$ in the test spectrum.
The purple curve is for NH and the blue curve is for IH. In generating those plots, we have calculated octant sensitivity for each values of true $\dcp$ and choose the
minimum $\chi^2$ for given value of true $\theta_{23}$. Thus the sensitivity reflected in the figure corresponds to the conservative values of the $\dcp$. 
From the figure we see that the octant sensitivity of HK is 
poor. It can only determine octant if the true value of $\theta_{23}$ lies between $40^\circ$ - $42^\circ$ for NH at $2 \sigma$ sensitivity and for 
IH almost the full parameter space is allowed at $2 \sigma$. 

Now let us discuss the CP sensitivity of the HK experiment. In the lower row of Fig.\,\ref{hk-atm-chi2}, we have plotted the CP violation discovery $\chi^2$ vs true $\dcp$ for 
five different true values
of $\theta_{23}$. Left panel is for NH and the right panel is for IH. The CP violation discovery potential of an experiment is defined by its capability to a distinguish a true value
of $\dcp$ from test value of $0^\circ$ or $180^\circ$. In all our plots we have minimized the $\chi^2$ with respect to sign($\Delta m^2_{\mbox{\scriptsize\rm eff}}$) and $\theta_{23}$
in the test spectrum.
From the figures we conclude that CP sensitivity in NH is higher as compared to 
the CP sensitivity in IH. For NH, $\theta_{23}=40^\circ$, $42^\circ$ and $44^\circ$
have almost the same CP sensitivity and one can have a $3 \sigma$ sensitivity at $\dcp=+90^\circ$ for these $\theta_{23}$ values. Then as $\theta_{23}$ increases, CP sensitivity decreases. 
On the other hand, in the case of IH for $\theta_{23} = 40^\circ$ and $42^\circ$
one can have a $2 \sigma$ CP sensitivity around $\dcp=\pm 90^\circ$ and the for the other values CP sensitivity deteriorates due to the presence
of degeneracy.

\begin{figure*}
\begin{tabular}{lr}
\includegraphics[width=0.6\textwidth]{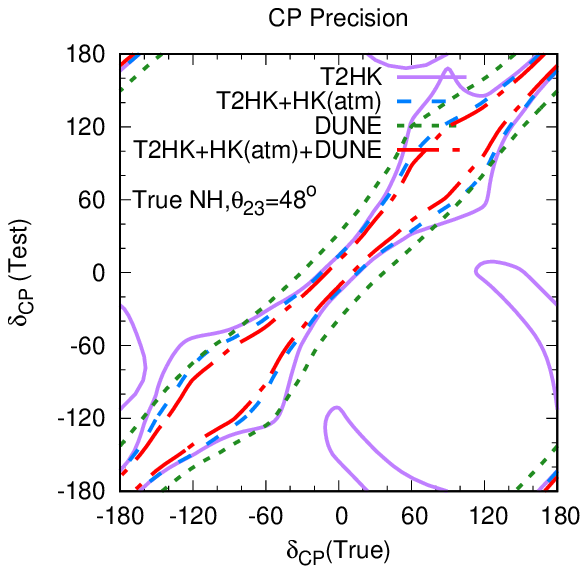}
\hspace*{-1.4in}
\includegraphics[width=0.6\textwidth]{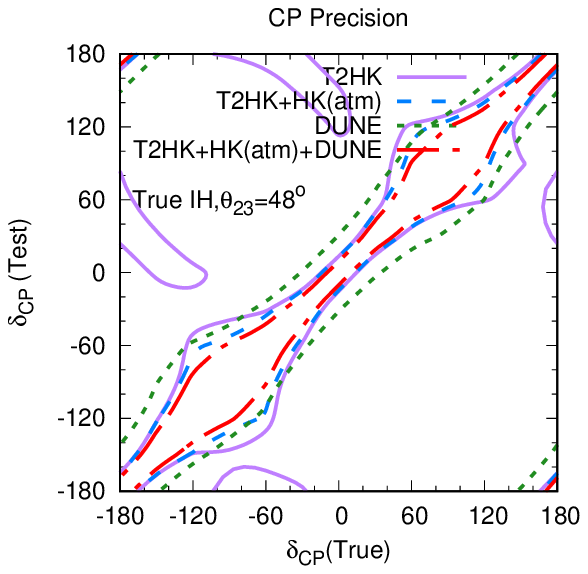} 
\end{tabular}
\caption{The correlation of the true and test
CP phases by individual experiment or by various combinations
of the experiments for $\theta_{23}=48^\circ$ at $3\sigma$ C.L.}
\label{cp_precision}
\end{figure*}

\subsection{Combined Sensitivity of T2HK, HK and DUNE}

In this section we will study the synergy between T2HK, HK and DUNE in determining the unknowns of neutrino oscillation and also study up to what confidence level these unknowns can be determined.

\subsubsection{Hierarchy}

In the upper row of Fig.\,\ref{hierarchy},
we plot
the significance
$\sqrt{\chi^2}$ at which hierarchy can be determined
as a function of true $\dcp$.
The left panel is for true NH and the right panel is for true IH.
These plots are similar as that of top left panel of Fig.\,\ref{hk-atm-chi2}.
As it is explained in Subsect.\,\ref{t2hk-dune},
in the case of $\dcp=-90^\circ$ with NH
and $\dcp=90^\circ$ with IH, T2HK can
resolve the mass hierarchy and the sensitivity
of T2HK is good, whereas
in the case of $\dcp=90^\circ$ with NH
and $\dcp=-90^\circ$ with IH, T2HK cannot
resolve the mass hierarchy by itself,
so the sensitivity of T2HK becomes poor.
If we combine T2HK and the Hyperkamiokande atmospheric
neutrino data, 
then T2HK+HK(atm)
can resolve the sign degeneracy at
5\,$\sigma$ C.L. for any value of $\dcp$.
Here it is important to note that for IH, in the region $-180^\circ < \dcp < 0^\circ$, both T2HK and HK has poor hierarchy sensitivity. In this parameter space, 
T2HK suffers from hierarchy degeneracy and
HK suffers from octant degeneracy. But when these experiments are combined there is a great enhancement in the sensitivity in that unfavorable values of $\dcp$. 
This reflects the synergy between these two
experiments which are essential to achieve a hierarchy sensitivity around $5 \sigma$ C.L. irrespective of the true value $\dcp$.
On the other hand, for
DUNE, the separation of the two mass hierarchies
are good and DUNE itself has above $5 \sigma$ C.L. sensitivity  to
the mass hierarchy for any value of $\dcp$.
If we combine T2HK, the HK atmospheric
neutrino data and DUNE, then
the significance of the mass hierarchy
becomes as large as
15\,$\sigma$ C.L., even for the unfavorable values of $\dcp$. This is a quite remarkable
result which shows the potential of these experiments to discover neutrino mass hierarchy.

\subsubsection{Octant}

In the middle panels of Fig.\,\ref{hierarchy}, we plot
the significance
$\sqrt{\chi^2}$ at which the wrong octant can be excluded
as a function of true $\theta_{23}$. The left panel is for true NH and the right panel is for true IH. 
These figures are similar as that of the top right panel of Fig.\,\ref{hk-atm-chi2} i.e., our results
corresponds to the conservative values of $\dcp$.
From the plots we see that the values of $\theta_{23}$ for which octant can be resolved at $5 \sigma$ C.L is almost the same for T2HK, T2HK+HK and DUNE. For these setups octant can be determined 
except $43^\circ < \theta_{23} < 49^\circ$ for both NH and IH at $5 \sigma$ C.L. However when all the three 
experiments are combined we see that there is a significant amount of increase in the octant sensitivity
for both the hierarchies. From the figure we see that T2HK+HK+DUNE can resolve octant except $43.5^\circ < \theta_{23} < 48^\circ$ for both the hierarchies at $5 \sigma$ C.L. 

\begin{figure*}
\begin{tabular}{lr}
\vspace{-0.2 in}
\includegraphics[width=0.5\textwidth]{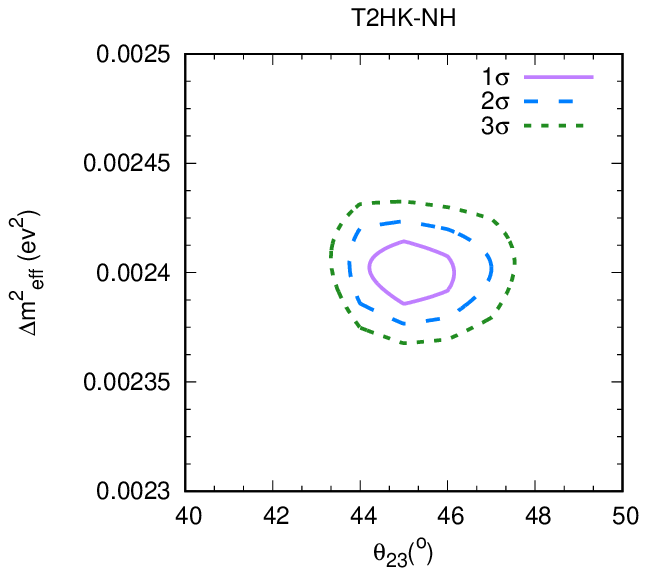}
\hspace*{-0.9in}
\includegraphics[width=0.5\textwidth]{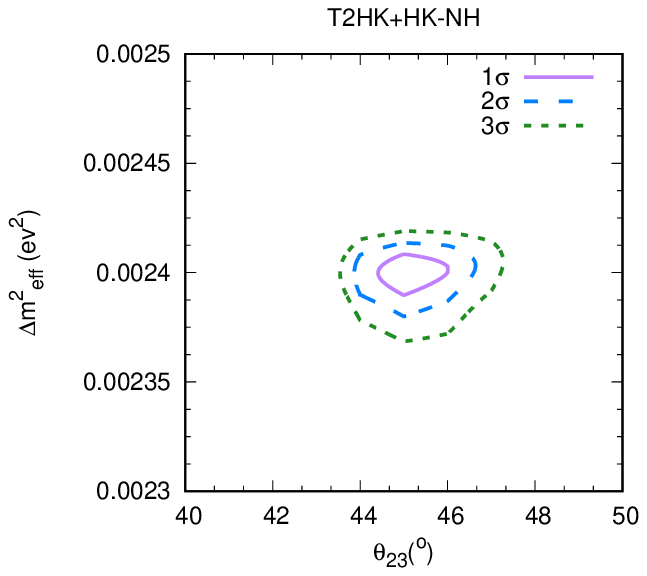} \\
\vspace{-0.2 in}
\includegraphics[width=0.5\textwidth]{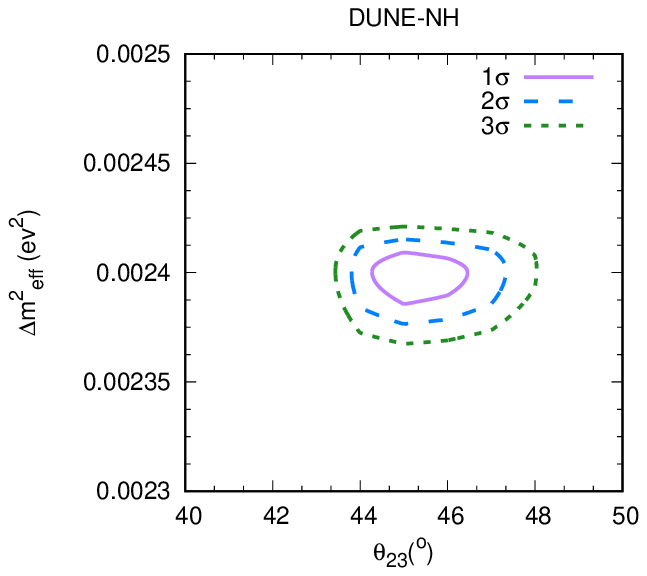}
\hspace*{-0.9in}
\includegraphics[width=0.5\textwidth]{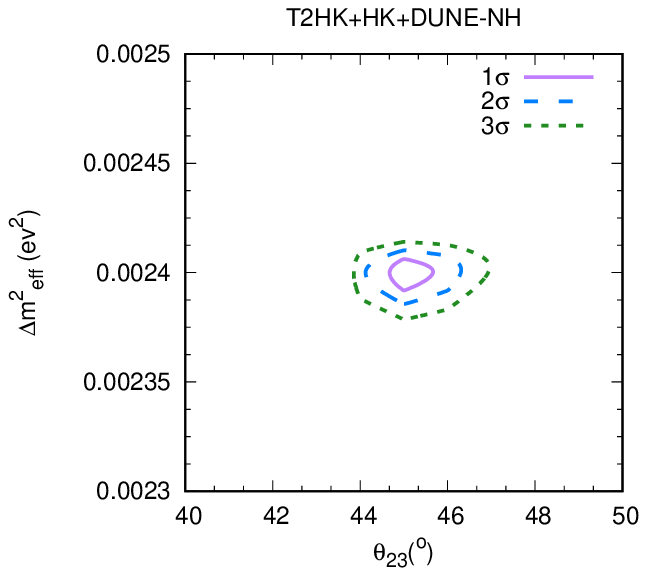} \\ 
\vspace{-0.2 in}
\includegraphics[width=0.5\textwidth]{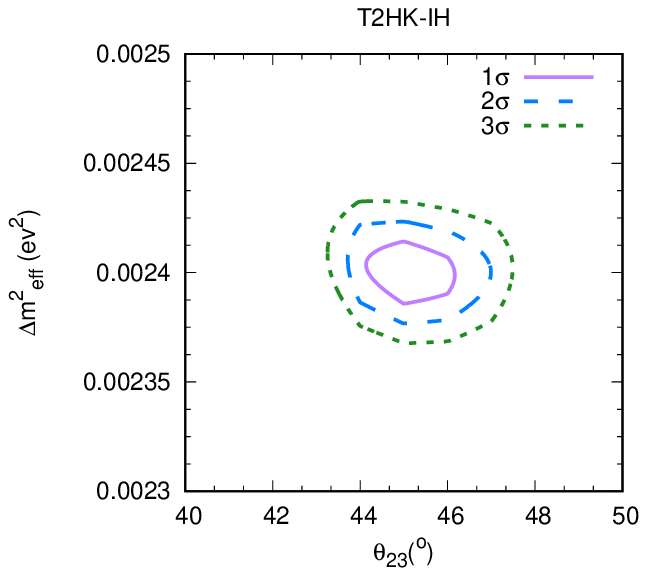}
\hspace*{-0.9in}
\includegraphics[width=0.5\textwidth]{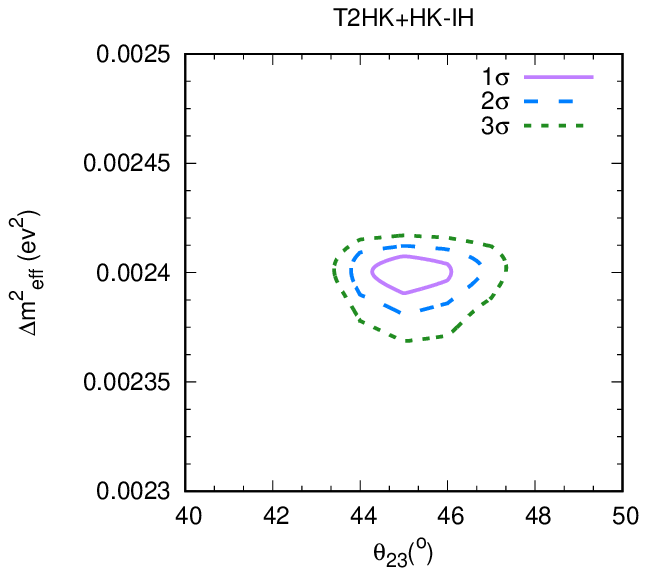} \\ 
\includegraphics[width=0.5\textwidth]{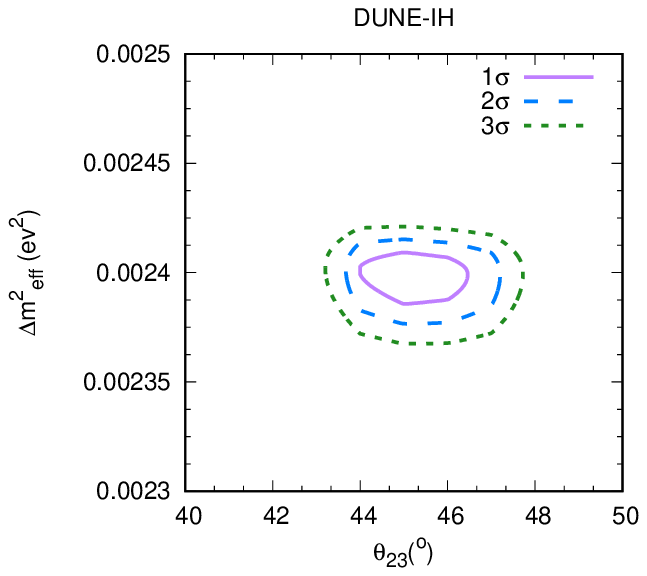}
\hspace*{-0.9in}
\includegraphics[width=0.5\textwidth]{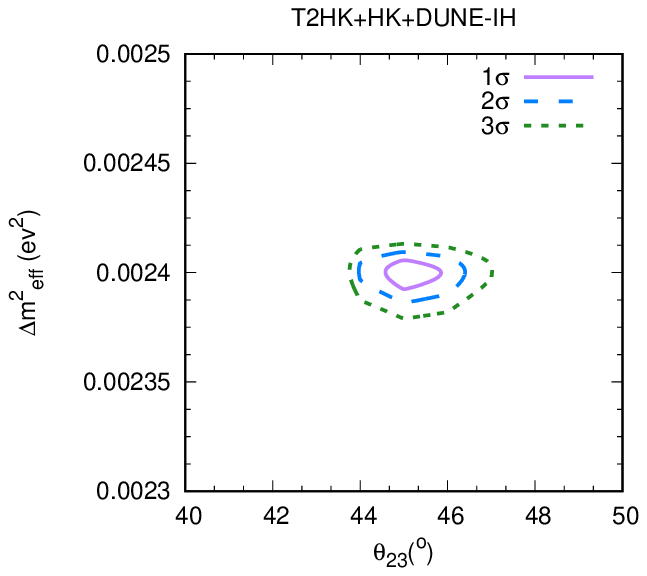} 
\end{tabular}
\caption{The correlation between
$\Delta m^2_{{\rm eff}}$ and
$\theta_{23}$ for various combinations
of the experiments.  Normal (inverted) hierarchy
is assumed in the first and second (third and fourth) row.}
\label{param_precision2}
\end{figure*}


\subsubsection{CP Violation}

In the lower panels of Fig.\,\ref{hierarchy} we plot the fraction of true $\dcp$ values for which CP violation can be discovered at a $5 \sigma$ confidence level as function of true $\theta_{23}$.
The left panel is for NH and the right panel is for IH. From the plots we see that as $\theta_{23}$ increases the percentage of true $\dcp$ decreases. 
For T2HK+HK, the $\dcp$ fraction varies from 60\% to
70\% and for T2HK+HK+DUNE, the $\dcp$ fraction varies from 75\% to 68\%. From these results we clearly understand that the combination of these three experiments have the capability to discover
CP violation at $5 \sigma$ C.L. for at least 68\% true values of $\dcp$. This is an incredible result since no other existing facilities can have such a reach to discover CP violation.

From these figures we also note that the sensitivities of DUNE and T2HK are similar. To understand this in Fig.\,\ref{cp}, 
we plot the CP violation discovery $\chi^2$ as a function of true $\dcp$ for
$\theta_{23}=48^\circ$ which is the present best-fit value of $\theta_{23}$.
As was discussed in Subsect.\,\ref{t2hk-dune},
the sensitivity of T2HK to CP is poor for
$\dcp\sim 90^\circ$ with NH and $\dcp\sim-90^\circ$ with IH. On the other hand due to large matter effect DUNE has almost equal sensitivity at $\pm 90^\circ$ for 
both the hierarchies. But due to very high statistics,
T2HK has very large CP sensitivity in the favorable values of $\dcp$ and this gives almost comparable coverage of true $\dcp$ values for which CP violation can be discovered at $5 \sigma$ C.L. 
It is also important to note that
if we combine the result of the Hyperkamiokande
atmospheric neutrino data to T2HK, then
sensitivity is good also for the unfavorable values of $\dcp$. This is because the hierarchy sensitivity of HK, removes the wrong hierarchy solutions of T2HK in the unfavorable parameter space.
We note in passing that synergy between T2K and the atmospheric neutrino
measurement at ICAL@INO for CP violation discovery was studied in Ref.\,\cite{Ghosh:2013zna}.
Furthermore, if we combine T2HK, the Hyperkamiokande
atmospheric neutrino and DUNE, then the sensitivity to
CP violation reaches around 10\,$\sigma$ C.L. for
$\dcp\sim-90^\circ$ for both NH and IH.

\begin{table*}
\begin{center}
\begin{tabular}{|c|c|c|c|c|}
\hline
Sensitivity   & T2HK & T2HK+HK(atm)  & DUNE & {\small T2HK+HK(atm)+DUNE}\\          
\hline
Hierarchy &    1   & 5    &  8  & 15    \\
Octant      &   $43^\circ - 49^\circ$   & $43^\circ - 49^\circ$    &  $43^\circ - 49^\circ$  & $43.5^\circ - 48^\circ$   \\
CP Violation  &    30  & 60     &  20  & 68    \\
\hline

\hline
\end{tabular}
\end{center}
\caption{Sensitivity for hierarchy, octant and CP at conservative values of the true parameter.
For hierarchy,  $\sqrt{\chi^2}$ is given.
For octant, the region for $\theta_{23}$ which
is allowed at $5 \sigma$ is given.
For CP violation, the fraction of $\dcp$ for which CPV can be discovered at $\chi^2=25$ is given in $\%$.
}
\label{tab:adequate} 
\end{table*}

\subsection{Precision of $\dcp$, $\theta_{23}$ and $\Delta m^2_{{\rm eff}}$}

In this section we study how much precisely the parameters $\dcp$, $\theta_{23}$ and $\Delta m^2_{{\rm eff}}$ can be measured by the set up under consideration.
In Fig.\,\ref{param_precision} we show
the $90\%$ C.L. precision contours in the test $\theta_{23}$ and test $\dcp$ plane
for various true values by individual experiments or by combinations
of the experiment. The true $\theta_{23}$ values are considered as $40^\circ$ to $50^\circ$ with a step of $2^\circ$ and the true $\dcp$ values are 
considered as $-180^\circ$ to $90^\circ$ with a step of $90^\circ$.
The upper panels are for NH and the lower panels are for IH. 
Because of the
difficulty of resolving the sign degeneracy
at T2HK, T2HK alone may lead to wrong
region for the test CP phase. 
This can be seen from the top left and bottom left panels of Fig.\,\ref{param_precision} which corresponds to the T2HK experiment. 
For true $\dcp=-180^\circ$, the wrong regions are seen around $-30^\circ$ ($+30^\circ$) for NH (IH). But in DUNE there are no wrong solutions at all (the top third and bottom third panels). 
Note that when we combine T2HK with the HK atmospheric
neutrino data, then the
fake region for the test CP phase disappears (the top second and bottom second panels).The precision is seen to be excellent when all the three experiments are combined (the top fourth and bottom fourth panel).

To understand the synergy between different experiments in improving the CP precision, 
in Fig.\,\ref{cp_precision}, we plot the $3\sigma$ contours in the $\dcp$ (true) vs $\dcp$ (test) plane for a fixed value of $\theta_{23}=48^\circ$. 
The left panel is for NH and the right panel is for IH. From the plot we see that, for T2HK there are wrong $\dcp$ solutions mainly in the region $0^\circ < \dcp < 180^\circ$ in NH and
$-180^\circ < \dcp < 0^\circ$ in IH. But when the HK is added with it, the wrong solutions completely disappears. But in the case of DUNE there are no wrong $\dcp$ solutions. 
It is also seen that the CP precision
is excellent when all the three experiments are combined. Here it is interesting to note that the precision of $\dcp$ is better at $\dcp=0^\circ$ as compared to $\dcp=\pm 90^\circ$.

Finally in Fig.\,\ref{param_precision2} we
plotted the $1 \sigma$, $2 \sigma$ and $3\sigma$ contours in the
$\Delta m^2_{{\rm eff}}$ - $\theta_{23}$ plane for various combinations
of the experiments. The true value is taken as $\theta_{23}=45^\circ$ and $|\Delta m_{{\rm eff}}^2| = 0.0024$ eV$^2$. In generating these plots we have kept $\dcp$ fixed at $-90^\circ$
in both true and test spectrum.
The upper panels are for NH and the lower panels are for IH.
From the plots we see that T2HK and DUNE have similar sensitivity.
If we combine
T2HK, the HK atmospheric
neutrino data and DUNE, then
the errors in $\Delta m^2_{{\rm eff}}$,
$\sin^2\theta_{23}$ and $\dcp$ become 0.3\%, 2\% and 20\%,
respectively.

\subsection{Precision of $\theta_{13}$}
In this paper we have discussed only the measurements of
$\Delta m^2_{{\rm eff}}$, $\theta_{23}$ and $\dcp$.
Let us discuss briefly whether the combination of T2HK+HK+DUNE
can improve the current precision on $\theta_{13}$
which is obtained by the reactor experiments.
It is known\,\cite{Yasuda:2004gu}
that if we fix the values of the both appearance
probabilities $P(\nu_\mu\to\nu_e)$ and
$P(\bar{\nu}_\mu\to\bar{\nu}_e)$, then it
gives us a quadratic curve in the
($\sin^22\theta_{13}$, $1/\sin^2\theta_{23}$) plane,
where a point sweeps the quadratic curve as $\dcp$ varies
from $0$ to $2\pi$.
The region of the quadratic curve in the
($\sin^22\theta_{13}$, $1/\sin^2\theta_{23}$) plane
is the necessary and sufficient condition which can
be obtained from the appearance
probabilities $P(\nu_\mu\to\nu_e)$ and
$P(\bar{\nu}_\mu\to\bar{\nu}_e)$ only.
To get information on $\theta_{13}$ from
this quadratic curve without the reactor data,
we need to combine the disappearance probabilities
$P(\nu_\mu\to\nu_\mu)$ and
$P(\bar{\nu}_\mu\to\bar{\nu}_\mu)$ with
the appearance probabilities.
Because the error $\delta(\sin^2\theta_{23})$
=$(1/4)\delta(\sin^22\theta_{23})/(1-\sin^22\theta_{23})^{1/2}$
is enhanced due to the singular Jacobian
factor at the maximal mixing\,\cite{Minakata:2004pg},
the uncertainty in the vertical coordinate
in the ($\sin^22\theta_{13}$, $1/\sin^2\theta_{23}$) plane
is expected to be large.
Notice that the uncertainty in the horizontal coordinate
$\sin^22\theta_{13}$ is not enhanced in the case of
the reactor measurements, because
$\sin^22\theta_{13}$ appears linearly in the disappearance probability
in the reactor experiments.
From this discussion, we expect that
it is difficult for the combination of T2HK+HK+DUNE
to give a precision on $\theta_{13}$
which is competitive with the current one
from the reactor experiments.

\section{Conclusion} 
\label{conclusion}

In this paper we have studied the sensitivity
of T2HK, HK and DUNE to mass hierarchy, octant of the mixing angle $\theta_{23}$ and $\dcp$.
The main results of our analysis are summarized in Table \ref{tab:adequate}.
Although it is difficult for T2HK to resolve
the sign degeneracy for unfavorable region
of the CP phase, when we combine it with
the atmospheric neutrino measurement at
Hyperkamiokande, we can determine the
mass hierarchy at 5\,$\sigma$ C.L. for any
value of $\dcp$.  We have also clarified
how the octant degeneracy occurs 
and why its behavior depends on
the mass hierarchy in the
HK atmospheric neutrino measurements.
On the other hand, DUNE can determine the
mass hierarchy at least at 8\,$\sigma$ C.L.
by itself.  Furthermore, if we combined
all of them, then the significance to
mass hierarchy is at least 15\,$\sigma$ C.L.
In our analysis we found out that the octant sensitivity of T2HK, T2HK+HK and DUNE are quite similar in ruling out the wrong octant at $5 \sigma$ C.L.
But for T2HK+HK+DUNE the increase in the octant sensitivity is significant.
For CP violation discovery we find that the combination T2HK+HK
can measure
CP violation at 8\,$\sigma$ C.L. for
$\dcp=\pm 90^\circ$ and for T2HK+HK+DUNE
the significance for
CP violation is around 10\,$\sigma$ C.L.
for $\dcp=\pm 90^\circ$.
It is also quite impressive that with the combination of all the three experiment CP violation can be established at $5 \sigma$ C.L for at least $68\%$ true values of $\dcp$.
In the combination of all these
experiments above, the precision in $\Delta m^2_{{\rm eff}}$,
$\theta_{23}$ and $\dcp$ is 0.3\%, 2\% and 20\%.
The precision in the first two parameters is improved
by one order of magnitude
compared with the current data.
We will be in the era of
precision measurements of neutrino
oscillation parameters, and
combination of Hyperkamiokande and DUNE
will play an important role in determination
of $\dcp$ as well as $\theta_{23}$.

\section*{Acknowledgement}

MG would like to thank Srubabati Goswami for useful discussions.
This work is supported by the ``Grant-in-Aid for Scientific Research of the Ministry of Education, 
Science and Culture, Japan", under 
Grants No. 25105009, No. 15K05058, No. 25105001 and No. 15K21734.

\appendix

\section{Octant degeneracy in the atmospheric neutrinos}
\label{appendixa}

\begin{figure*}
\begin{tabular}{lr}
\hspace{-0.05 in}
\includegraphics[width=0.45\textwidth]{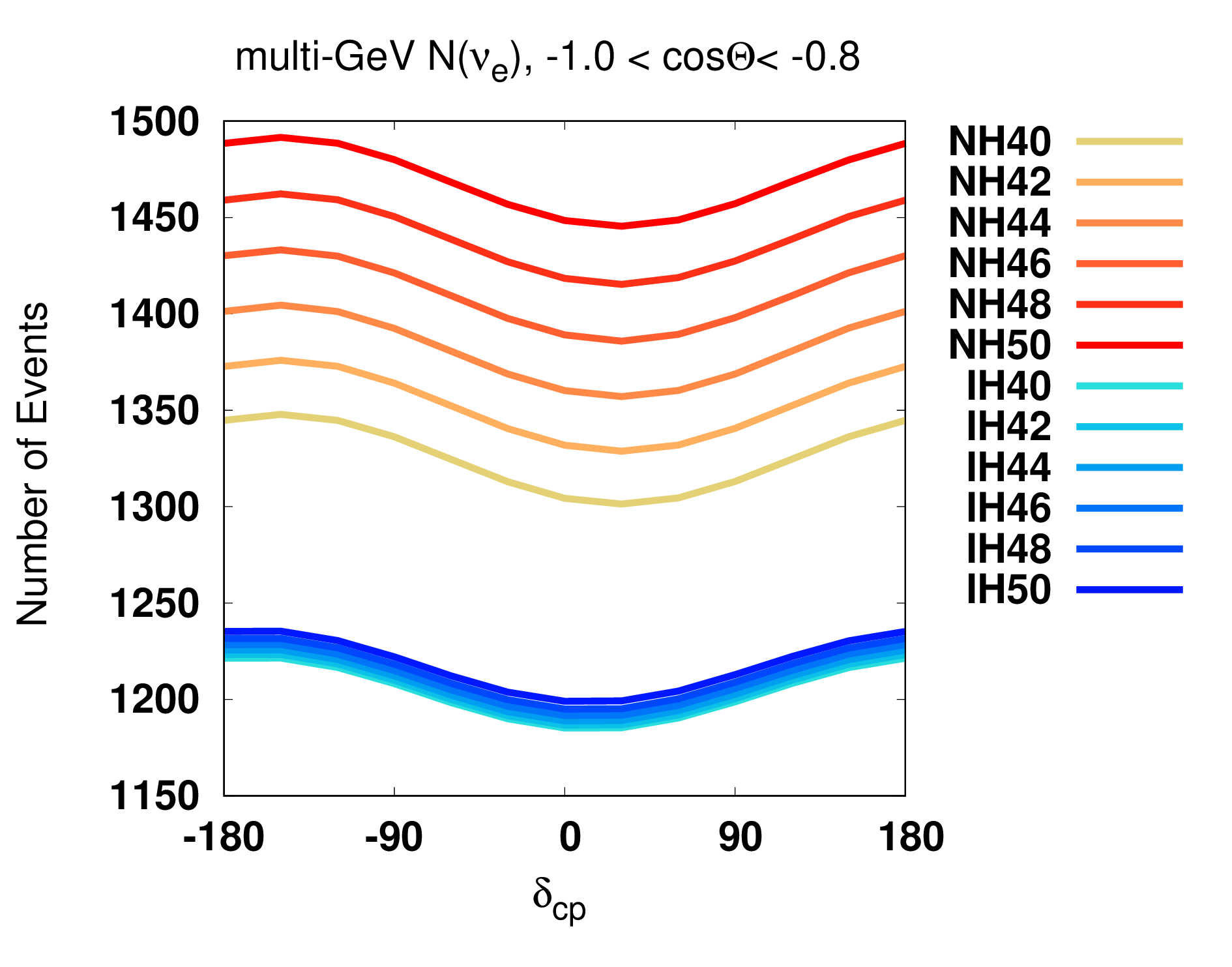}
\hspace{-0.05 in}
\includegraphics[width=0.45\textwidth]{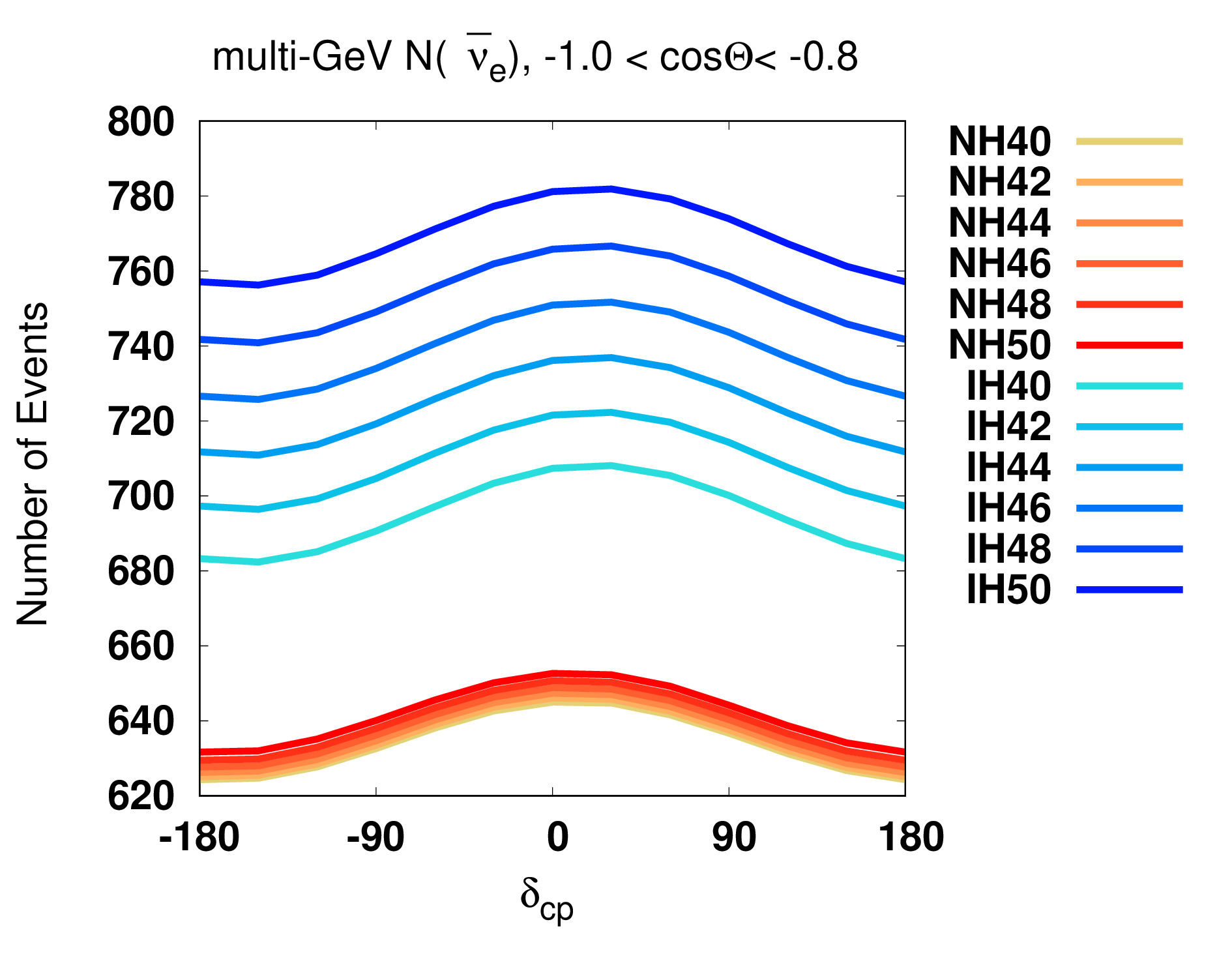} \\
\hspace{-0.05 in}
\includegraphics[width=0.45\textwidth]{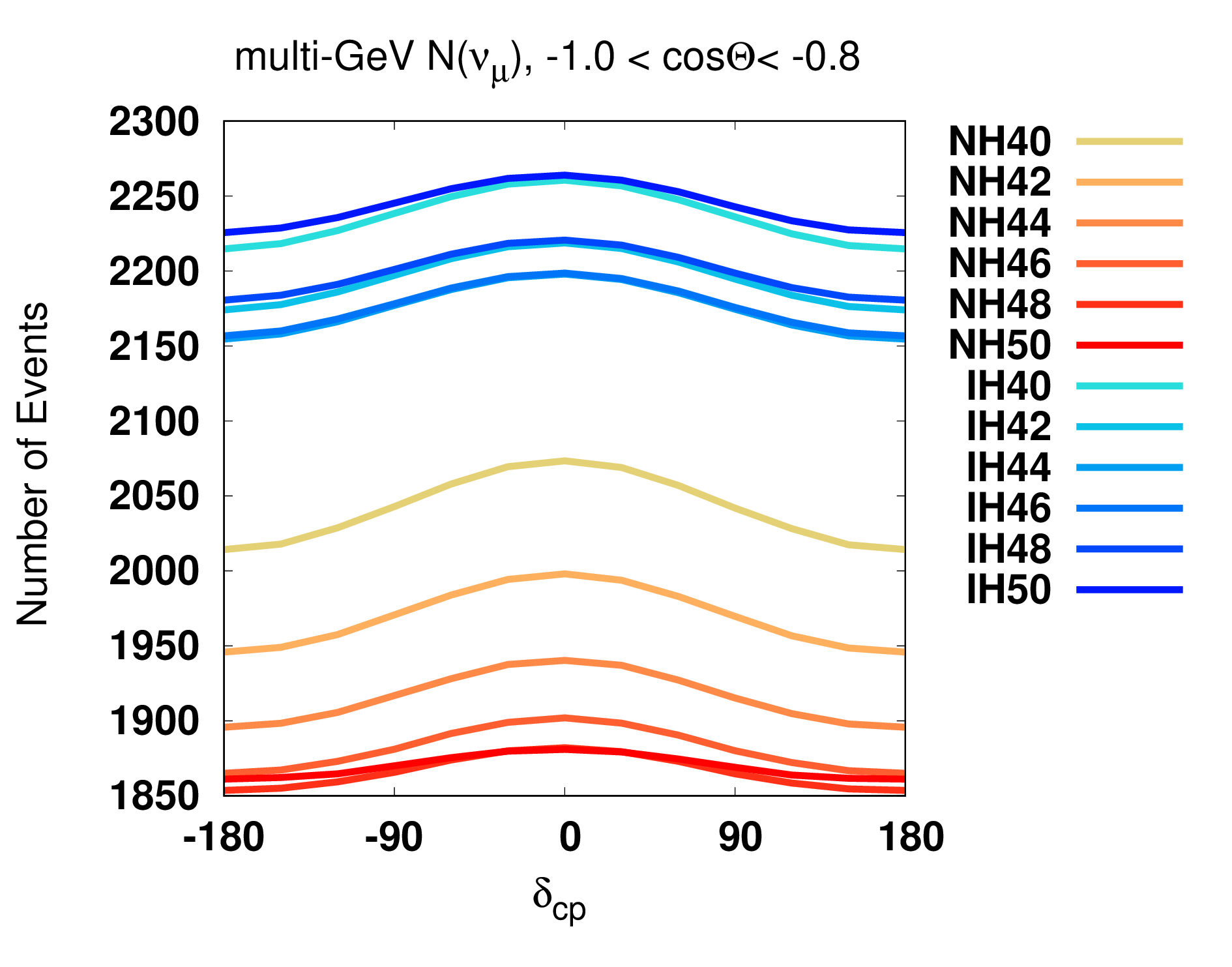}
\hspace{-0.05 in}
\includegraphics[width=0.45\textwidth]{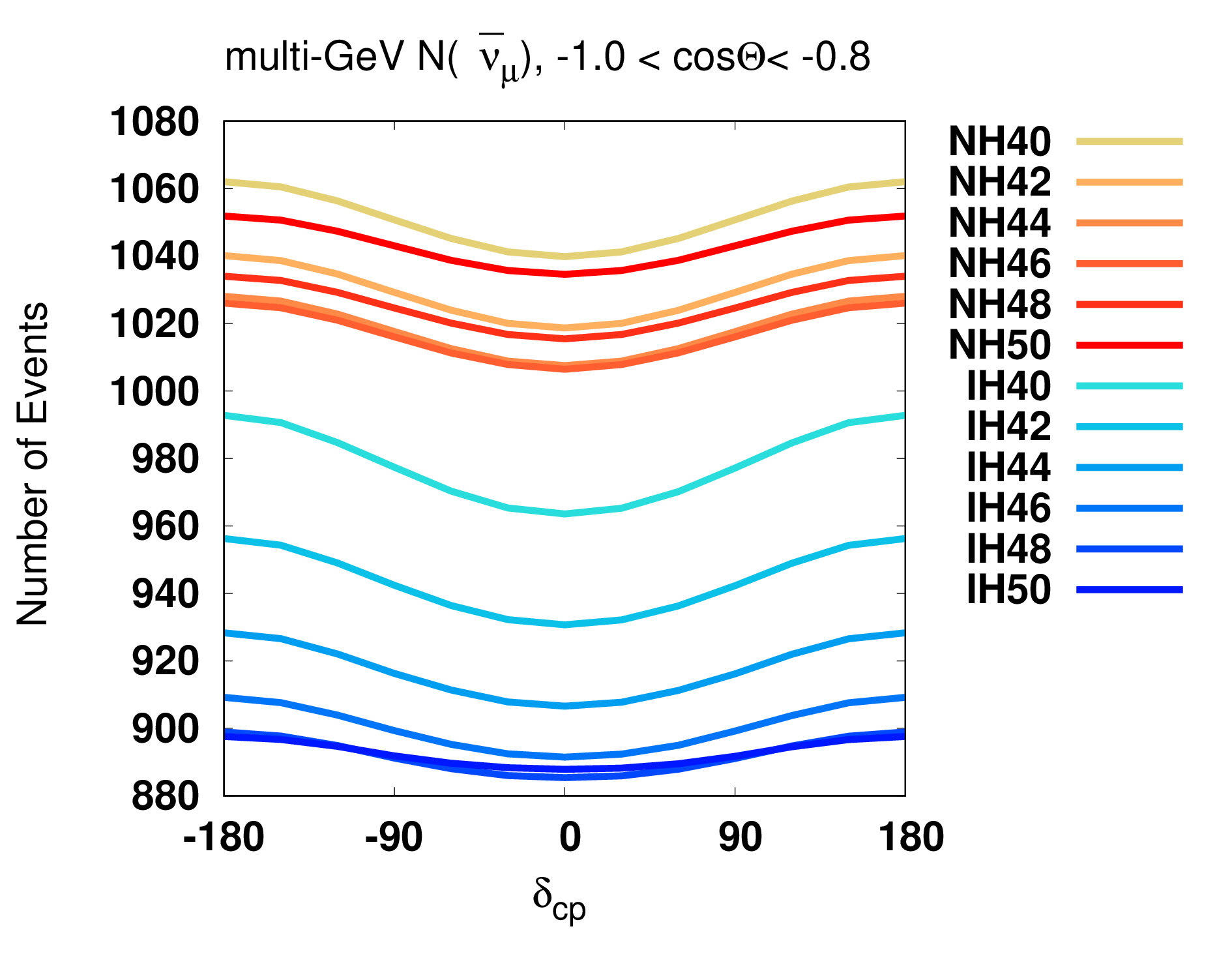}
\end{tabular}
\caption{The numbers of the multi-GeV events for the zenith angle
$-1<\cos\Theta<-0.8$.
Left panels: neutrino events;\quad
Right panel: antineutrino events.}
\label{e1}
\end{figure*}

\begin{figure*}
\begin{tabular}{lr}
\hspace{-0.05 in}
\includegraphics[width=0.45\textwidth]{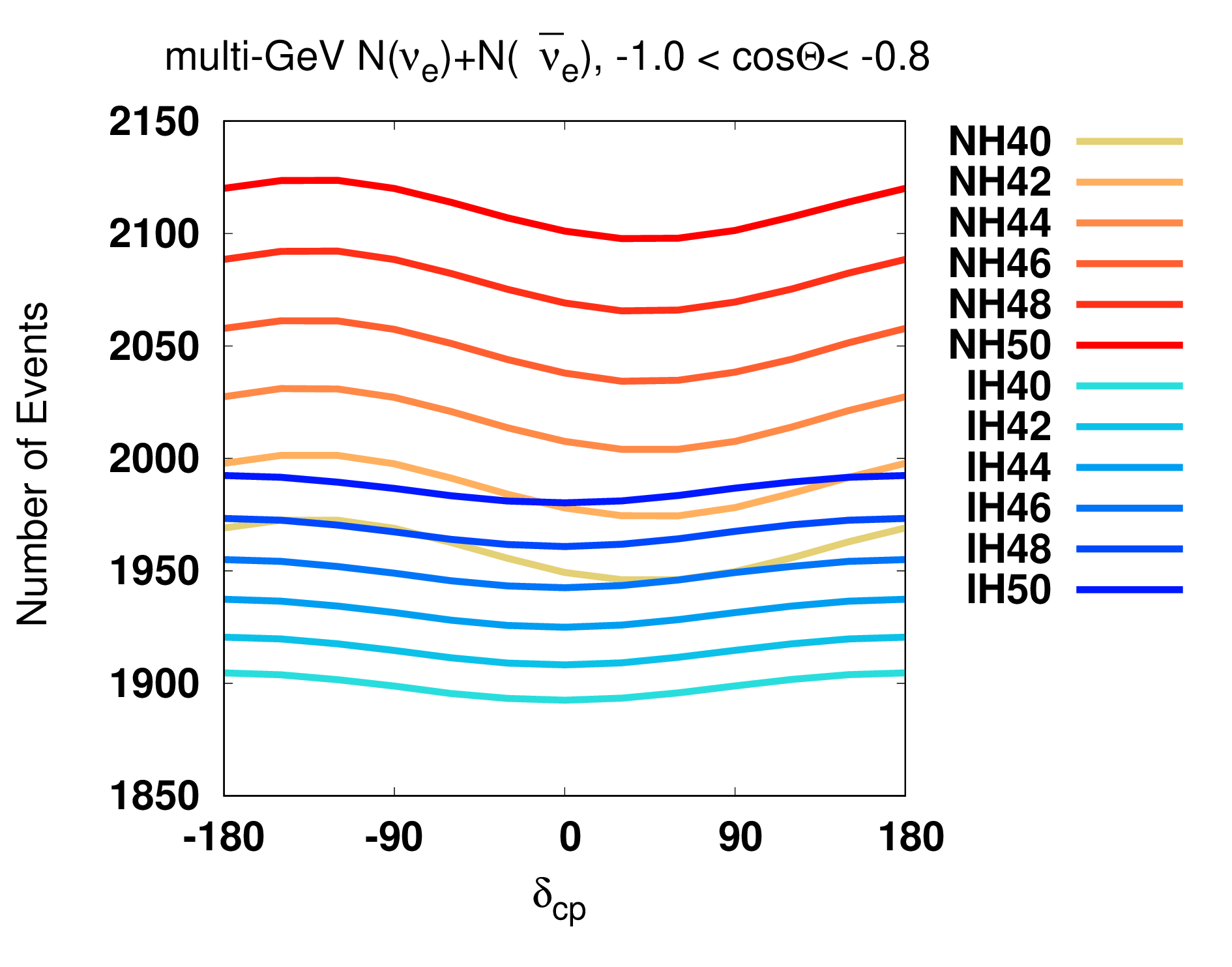}
\hspace{-0.05 in}
\includegraphics[width=0.45\textwidth]{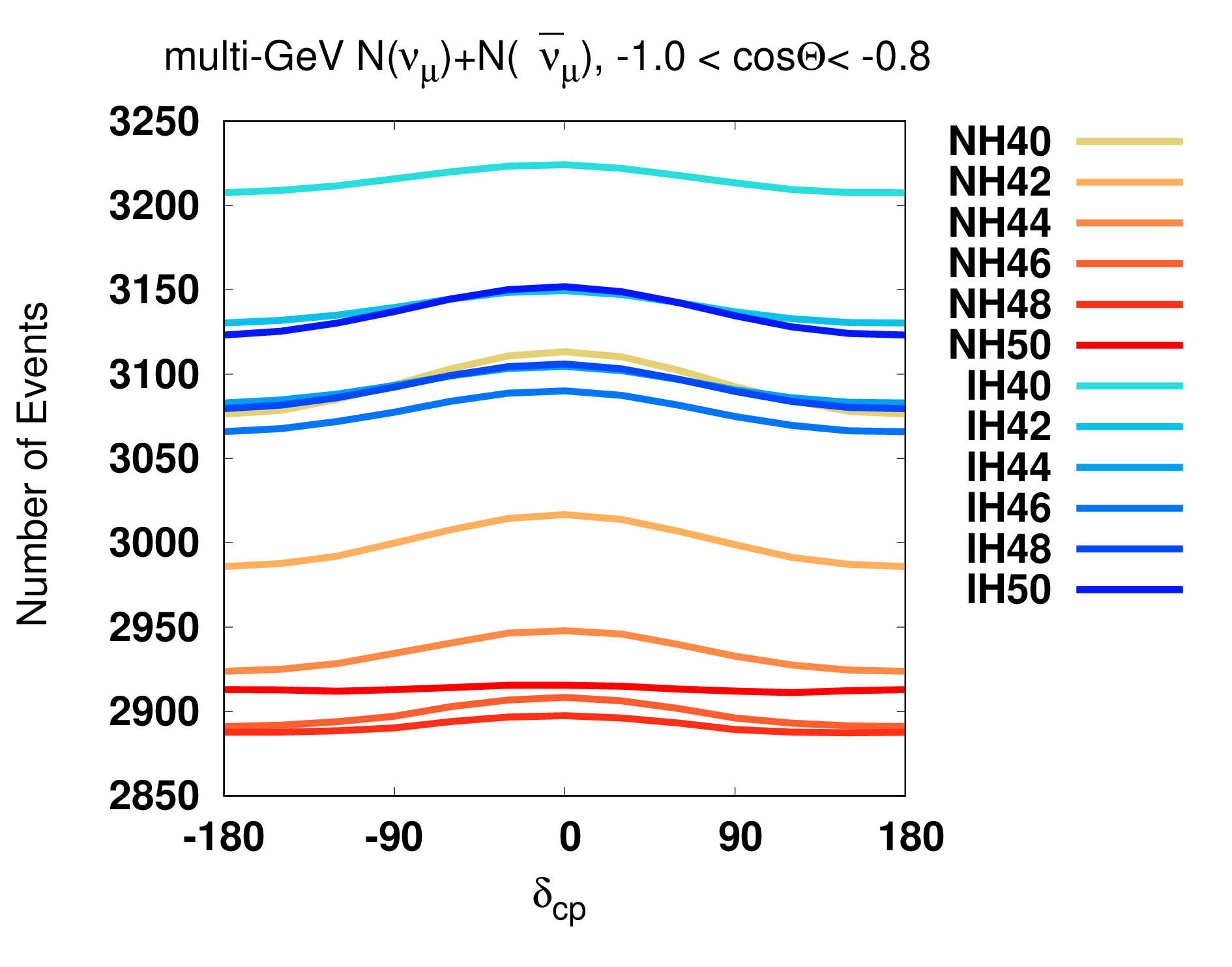}
\end{tabular}
\caption{The numbers of the multi-GeV events for the zenith angle
$-1<\cos\Theta<-0.8$.
Left panel: $N(\nu_e\to\nu_e)+N(\nu_\mu\to\nu_e)+N(\bar{\nu}_e\to\bar{\nu}_e)+N(\bar{\nu}_\mu\to\bar{\nu}_e)$;
Right panel: $N(\nu_e\to\nu_\mu)+N(\nu_\mu\to\nu_\mu)+N(\bar{\nu}_e\to\bar{\nu}_\mu)+N(\bar{\nu}_\mu\to\bar{\nu}_\mu)$}
\label{tot1}
\end{figure*}

In this appendix, we will discuss the octant degeneracy
in the atmospheric neutrino measurements and we will clarify
the reason why the significance of the mass hierarchy
differs depending on the true mass hierarchy.

The atmospheric neutrinos which can be directly measured
are e-like and $\mu$-like events and its sensitivity
to the mass hierarchy can be expressed by $\chi^2$
which is defined as\footnote{
In this appendix, for simplicity we ignore
the systematic errors, since our discussions are
only qualitative to understand parameter degeneracy in the
atmospheric neutrino measurements.}

\begin{eqnarray}
&{\ }&\hspace{-20mm}  \chi^2 
  = \sum_{j} \sum_{\beta = e, \mu}\left[
  \sum_{\alpha = e, \mu} \left\{N_j^{\alpha \beta}(WH)
  + N_j^{\bar{\alpha}\bar{\beta}}(WH)
- N_j^{\alpha \beta}(TH)
- N_j^{\bar{\alpha}\bar{\beta}}(TH)
  \right\}
\right]^2
\nonumber\\
&{\ }&\hspace{1mm}
\times\left[\sum_{\alpha = e, \mu} \left\{
N_j^{\alpha \beta}(TH)
+N_j^{\bar{\alpha}\bar{\beta}}(TH)
\right\}\right]^{-1}
\label{sk}
\end{eqnarray}
where we have introduced the
simplified notations
\begin{eqnarray}
&{\ }&\hspace{-5mm}
N_j^{\alpha\beta}(MH)
\equiv F_0(\nu_\alpha)\,
P(\nu_\alpha\to\nu_\beta, MH)
\,\sigma(\nu_\beta)
\nonumber\\
&{\ }&\hspace{-5mm}
N_j^{\bar{\alpha}\bar{\beta}}(MH)
\equiv F_0(\bar{\nu}_\alpha)\,
P(\bar{\nu}_\alpha\to\bar{\nu}_\beta, MH)
\,\sigma(\bar{\nu}_\beta)\,.
\nonumber
\end{eqnarray}
Here $F_0(\nu_\alpha)$ is the original
flux of $\nu_\alpha$ before oscillations,
$P(\nu_\alpha\to\nu_\beta, MH)$ is the
oscillation probability with a given
Mass Hierarchy (either the True Hierarchy
$MH = TH$ or the Wrong Hierarchy $MH = WH$),
$\sigma(\nu_\beta)$ ($\sigma(\bar{\nu}_\beta)$)
is the cross section
for $\nu_\beta$ ($\bar{\nu}_\beta$) to produce the charged lepton
$\ell^-_\beta$ ($\ell^+_\beta$),
and $j$ stands for the index for the zenith angle
bin ($(j-6)/5 < \cos\Theta < (j-5)/5$).

\begin{table*}
\begin{center}
\begin{tabular}{|c|c|c|c|c|c|c|c|c|}
\hline
MH & $N_1^{\mu e}$ & $N_1^{\bar{\mu}\bar{e}}$  & $N_1^{ee}$ & $N_1^{\bar{e}\bar{e}}$
& $N_1^{e\mu}$ & $N_1^{\bar{e}\bar{\mu}}$ & $N_1^{\mu\mu}$ & $N_1^{\bar{\mu}\bar{\mu}}$\\          
\hline
NH& 498 $\pm$ 108 & 66 $\pm$ 15 & 898 $\pm$ 1 & 573 $\pm$ 9$\times 10^{-2}$
& 11 $\pm$ 3 & 2 $\pm$ 5$\times 10^{-4}$ & 1970 $\pm$ 130 & 1040 $\pm$ 35\\
\hline
IH& 118 $\pm$ 26 & 259 $\pm$ 56 & 1093 $\pm$ 1 & 474 $\pm$ 9$\times 10^{-1}$
& 41 $\pm$ 10 & 3 $\pm$ 5$\times 10^{-3}$ & 2180 $\pm$ 70 & 950 $\pm$ 65\\
\hline
\end{tabular}
\end{center}
\caption{The numbers of events for the zenith bin
$j=1~(-1.0 < \cos\Theta < -0.8)$ with variations in
$\theta_{23}$ and $\delta$ for each mass hierarchy.
}
\label{tab:eventnumbers} 
\end{table*}

To see how degeneracy occurs, let us take a look,
for simplicity, at the
number of events for the zenith bin
$j=1~(-1.0 < \cos\Theta < -0.8)$,
in which the matter effect is expected to be
important.
Assuming that data is taken at Hyperkamiokande for
2000 days with 0.56 Mton fiducial volume,
the numbers of events
for each channel are estimated and are given in Table \ref{tab:eventnumbers}.
From these numbers of events
we can understand that the
main contribution for the
$\mu$-like events comes from
the $\nu_\mu\to\nu_\mu$ channel,
and that the
main contribution for the
$e$-like events comes from
the $\nu_\mu\to\nu_e$ channel.
For NH (IH) we have more events
in the neutrino (antineutrino) mode.
To understand
these results qualitatively,
we have to take into account a few facts.
First of all, in the case of the normal (inverted)
hierarchy, the oscillation probability
$P(\nu_\mu\to\nu_e)$ ($P(\bar{\nu}_\mu\to\bar{\nu}_e)$)
is enhanced due to the matter effect.
Secondly, around the neutrino
energy region $E\sim 10$GeV, the ratio
of the atmospheric neutrinos for the zenith bin
$j=1~(-1.0 < \cos\Theta < -0.8)$ is\,\cite{Honda:2015fha}
$F(\nu_\mu)$:$F(\bar{\nu}_\mu)$:$F(\nu_e$):$F(\bar{\nu}_e)$
$\simeq$ 40:40:10:7.  The reason
that the $\mu/e$ ratio is high at $E\sim 10$GeV is
because muons do not have enough time
to decay to produce $e^\pm$ and $\nu_e$ or
$\bar{\nu}_e$ at such a high energy.
Thirdly, the ratio of the cross sections for
$\nu_\alpha$ and $\bar{\nu}_\alpha$ is approximately 2:1.
Thus we have more numbers of events
$N(\nu_\mu\to\nu_\alpha)$,
rather than $N(\bar{\nu}_\mu\to\bar{\nu}_\alpha)$,
$N(\nu_e\to\nu_\alpha)$ and
$N(\bar{\nu}_e\to\bar{\nu}_\alpha)$,
and the mass hierarchy should be
the normal hierarchy to have enhancement
for $N(\nu_\mu\to\nu_e)$.

In Fig.\,\ref{e1}
we show the numbers of the multi-GeV events
($e$-like, $\mu$-like)
for the zenith angle $-1 < \cos\Theta < -0.8$, assuming that
we can separate the neutrino and antineutrino
modes.  The left (right) panels are for
the neutrino (antineutrino) modes.
Because of the enhancement due to the matter effect,
a remarkable dependence on $\theta_{23}$ as well as
separation between the two mass hierarchies can be seen
for NH (IH) in the neutrino (antineutrino) mode.
These features can be seen for the 
zenith angle region $-1 < \cos\Theta \lesssim -0.4$.
For the zenith angle region $\cos\Theta \gtrsim -0.4$,
the matter effect is not so dramatic, and we do not have
the distinction between the two mass hierarchies.

The qualitative behaviors
in Fig.\,\ref{e1}
can be roughly understood from the
analytic expressions of the
oscillation probabilities.
Using the formalism
by Kimura-Takamura-Yokomakura\,\cite{Kimura:2002hb,Kimura:2002wd}
on the exact analytic expression
for the oscillation probability in matter
with constant density,
it can be shown to first order
in $|\Delta m^2_{21}/\Delta m^2_{31}|$
and to arbitrary order in $\theta_{13}$
that the appearance
and disappearance probabilities
satisfy the following behaviors:

\begin{eqnarray}
&{\ }&\hspace{-15mm}
P(\nu_\mu\to\nu_\mu; \theta_{23}^\prime)
-P(\nu_\mu\to\nu_\mu; \theta_{23})
\nonumber \\
&{\ }&\hspace{-20mm}
\simeq  (\sin^22\theta_{23}^\prime-\sin^22\theta_{23})
\left\{\cos^2\tilde{\theta}_{13}
\sin^2
\left(\frac{\Lambda_+ L}{2}
\right)
+\sin^2\tilde{\theta}_{13}
\sin^2
\left(\frac{\Lambda_- L}{2}
\right)
\right\}
\nonumber \\
&{\ }&\hspace{-20mm}
-(\sin^4\theta_{23}^\prime-\sin^4\theta_{23})
\sin^22\tilde{\theta}_{13}
\sin^2\left(
\frac{\Delta\tilde{E}_{31}L}{2}
\right)\,,
\label{pmmdel23}\\
&{\ }&\hspace{-16mm}
  P(\nu_\mu\to\nu_e; \theta_{23}^\prime)
-P(\nu_\mu\to\nu_e; \theta_{23})
\simeq (\sin^2\theta_{23}^\prime-\sin^2\theta_{23})
\sin^2\left(
\frac{\Delta\tilde{E}_{31}L}{2}
\right)\,,
\label{pmedel23}
\end{eqnarray}
where we have introduced the following notations
($G_F$ is the Fermi coupling constant
and $N_e$ is the density of electrons):
\begin{eqnarray}
\Delta E_{jk}&\equiv&\frac{\Delta m^2_{jk}}{2E}
\equiv\frac{m^2_j-m^2_k}{2E},~
A\equiv\sqrt{2}G_FN_e, ~
\nonumber \\
\Delta \tilde{E}_{31}&\equiv&
\left\{
(\Delta E_{31}\cos2\theta_{13}-A)^2\right.
\left.+(\Delta E_{31}\sin2\theta_{13})^2
\right\}^{1/2}, ~ \nonumber \\
\Lambda_\pm &\equiv&
\frac{\Delta E_{31}+A\pm\Delta \tilde{E}_{31}}{2}, ~
|\tilde{U}_{\mu 1}|^2
\simeq
\frac{\Delta E_{31}(\Lambda_+-\Delta E_{31})}
{2\Delta \tilde{E}_{31}\Lambda_-}, ~
\nonumber \\
|\tilde{U}_{\mu 3}|^2
&\simeq&
\frac{\Delta E_{31}(\Delta E_{31}-\Lambda_-)}
{2\Delta \tilde{E}_{31}\Lambda_+}, ~
\tan 2\tilde{\theta}_{13}
\equiv\frac{\Delta E_{31}\sin2\theta_{13}}{\Delta E_{31}\cos2\theta_{13}-A}
\nonumber
\end{eqnarray}

For example, near the resonance where
the $\tilde{\theta}_{13}$ is close to $45^\circ$,
from Eq.\,(\ref{pmmdel23}) we see in the
case of NH that
the $\sin^4\theta_{23}$ terms becomes
dominant, and the behavior is consistent
with the bottom panels of Fig.\,\ref{e1}.
From the plot we see that
the behaviors of the neutrino events
are opposite to that of the
antineutrino events.
If one had a charge identification
of the events, then one could resolve
the wrong hierarchy - wrong octant
degeneracy.

In Fig.\,\ref{tot1}
we show the numbers of the multi-GeV events
($e$-like, $\mu$-like),
in which the neutrino and antineutrino
events are combined together, as is done
in most of the data of water \cnv\ detectors.
From this plot it is easy to see
that the NH - LO solution is
confused with the IH - HO
solution for both the $e$-like
and $\mu$-like events.
Since the variation of the numbers
of events with respect to $\theta_{23}$
is smaller in IH than in NH,
the minimum value of hierarchy $\chi^2$
is expected to be larger in the
case of $TH = NH$ than in the case
of $TH = IH$.
This implies that
the significance of the mass hierarchy
is larger in the
case of $TH = NH$ than in the case
of $TH = IH$.

\bibliography{T2HK_DUNEv2}

\end{document}